\documentclass[12pt]{article}
\usepackage[margin=1in]{geometry}  % set the margins to 1in on all sides
\usepackage[singlespacing]{setspace}
\usepackage{epsfig}
\usepackage{graphics,graphicx,color}
\usepackage{amssymb,amsfonts,amsthm,amsmath}
\usepackage{authblk}
\usepackage{multirow}
\usepackage{mathrsfs}
\usepackage{enumerate}
\usepackage{natbib}
\usepackage{float}
\usepackage{algorithm}
\usepackage{algpseudocode}
\usepackage{subcaption} 
\usepackage{url}
\usepackage[colorlinks=true, linkcolor=blue, citecolor=blue]{hyperref}
\usepackage{bm}
\usepackage{comment}
\usepackage{siunitx}
\usepackage{xcolor}

\usepackage{lscape}
\usepackage{mathtools}
\usepackage{stackengine} 
\usepackage{indentfirst}
 
\usepackage{array,arydshln}

\newtheorem{defn}{Definition}[section]
\newtheorem{exam}{Example}[section]

\newtheorem{thm}{Theorem}[section]

\DeclareMathOperator*{\argmin}{arg\,min}

\newcommand{\bzero}{0}

\newcommand{\bbeta}{\mbox{\boldmath${\beta}$} }

\DeclarePairedDelimiter\floor{\lfloor}{\rfloor}

\newcommand\oast{\stackMath\mathbin{\stackinset{c}{0ex}{c}{0ex}{\ast}{\bigcirc}}}

\title{Orthogonal Arrays: A Review} 

\author[1]{C. Devon Lin}
\author[2]{John Stufken}%  

\affil[1]{Department of Mathematics and Statistics, Queen's University}
\affil[2]{Department of Statistics, College of Engineering and Computing, George Mason University}
 
\begin{document}

\date{}
\maketitle
\begin{abstract}
 Orthogonal arrays are arguably one of the most fascinating and important statistical tools for efficient data collection. They have a simple, natural definition, desirable properties when used as fractional factorials, and a rich and beautiful mathematical theory. Their connections with combinatorics, finite fields, geometry, and error-correcting codes are profound. Orthogonal arrays have been widely used in agriculture, engineering, manufacturing, and high-technology industries for quality and productivity improvement experiments. In recent years, they have drawn rapidly growing interest from various fields such as computer experiments, integration, visualization, optimization, big data, machine learning/artificial intelligence through successful applications in those fields. We review the fundamental concepts and statistical properties and report recent developments. Discussions of recent applications and connections with various fields are presented.  

\end{abstract}

\section{Definition and History} \label{sec:definition}

Orthogonal arrays have applications in many areas and have proven to be a fascinating and rich subject for research. Statisticians, mathematicians and other researchers have studied orthogonal arrays since their introduction by C.\ R.\ Rao in a series of seminar papers \citep{rao1946hypercubes,rao1947factorial,rao1949class}. In the field of design of experiments they are used to determine the settings of factors for conducting experiments, so that effects of these factors on a response variable can be explored simultaneously. The possible settings for these factors are called {\em levels}. Let $S$ be a set of $s$ levels, conventionally denoted by $0,1,\ldots,s-1$. Formally, we can define an orthogonal array as follows. 

\begin{defn}\label{def:OA}
An $N \times k$ array $A$ with entries from $S$ is said to be an {\em orthogonal array} with $s$ levels, strength $t$, and index $\lambda$ if every $N \times t$ subarray of $A$ contains each $t$-tuple based on $S$ exactly $\lambda$ times as a row.    
\end{defn} 

We use $OA(N,s^k,t)$ to denote such an orthogonal array. Note that $\lambda$ is not included in the notation because it can be derived as $\lambda = N/s^t$. An $OA(N,s^k,t)$ can be used to determine $N$ input settings or level combinations for $k$ factors each with $s$ levels. Each row of an $OA(N,s^k,t)$ is called a {\em run} while each column corresponds to the setting of a factor or input variable. 

\begin{exam}
Table \ref{table:oa_8_4} lists an $OA(8,2^4,3)$, an orthogonal array with two levels,  strength three, and index unity. It has eight runs and it is for four factors. It has the property that every subarray consisting of its three distinct columns   contains all the eight level combinations exactly once.

\begin{table}[!h]
\begin{center}
\caption{An $OA(8,2^4,3)$}
\begin{tabular}{cccc}
\multicolumn{4}{c}{}\\
0 & 0 & 0 & 0 \\
0 & 0 & 1 & 1 \\
0 & 1 & 0 & 1 \\
0 & 1 & 1 & 0 \\
1 & 0 & 0 & 1 \\
1 & 0 & 1 & 0 \\
1 & 1 & 0 & 0 \\
1 & 1 & 1 & 1 \\
\end{tabular}\label{table:oa_8_4} 
\end{center}
\end{table}
\end{exam}

When all factors have the same number of levels, the arrays are often referred to as ``fixed-level'' or ``pure-level'' or {\em symmetrical} orthogonal arrays. In some experiments, different factors are allowed to have different numbers of levels, and this leads to a class of arrays named {\em mixed} (or {\em asymmetrical}) orthogonal arrays. They began to receive more attention in the early 1960s, especially through the consideration of asymmetrical orthogonal main effects plans in \cite{addelman1962orthogonal}. \cite{rao1973problems} discussed, among others, the notion of asymmetrical orthogonal arrays. Significant advances have been made during the past decades on their constructions 
\citep{wu1991balanced, wang1991approach, hedayat1992asymmetrical, sitter1993balanced, decock2000finding,suen2001asymmetric,pang2021construction}, resulting in wider applications. Formally, mixed (or asymmetrical) arrays can be defined as follows. 

\begin{defn}\label{def:MOA}
A mixed orthogonal array $OA(N, s_1^{k_1}s_2^{k_2}\cdots s_v^{k_v},t)$ is an $N \times k$ array where $k = k_1+k_2 + \cdots + k_v$ is the total number of factors, in which the first $k_1$ factors have $s_1$ levels, the next $k_2$ factors have $s_2$ levels, and so on, with the property that in any $N\times t$ subarray every possible $t$-tuple occurs an equal number of times as a row. 
\end{defn}

\begin{exam}
Table \ref{table:moa_12_5} lists an $OA(12,2^43^1,2)$, a mixed orthogonal array of strength two, with 12 runs, with the first four factors at two levels and the fifth at three levels. 

\begin{table}[!h]
\begin{center}
\caption{A mixed orthogonal array $OA(12,2^43^1,2)$ (transposed) }
\begin{tabular}{cccccccccccccc}
\multicolumn{12}{c}{}\\
0 & 0 & 1 & 1 & 0 & 0 & 1 & 1 & 0 & 0 & 1 & 1\\
0 & 1 & 0 & 1 & 0 & 1 & 0 & 1 & 0 & 1 & 0 & 1 \\
0 & 0 & 1 & 1 & 1 & 1 & 0 & 0 & 1 & 0 & 0 & 1\\
0 & 1 & 0 & 1 & 1 & 0 & 0 & 1 & 1 & 0 & 1 & 0 \\
0 & 0 & 0 & 0 & 1 & 1 & 1 & 1 & 2 & 2 & 2 & 2 \\ 
\end{tabular}\label{table:moa_12_5}
\end{center}
\end{table}
\end{exam}

% brief history of OAs 
The term orthogonal arrays was first introduced by \cite{bush1950}, although the concept was previously described in Rao's 1943 master thesis and three papers \citep{rao1946hypercubes, rao1947factorial, rao1949class}, first for the special case of what was called a {\em hypercube of strength $d$} \citep{rao1946hypercubes} and then for general orthogonal arrays. Orthogonal arrays have since found applications in various fields, including as fractional factorials for agricultural, medical, industrial, and other experiments. They are particularly useful in scenarios where the number of input variables is relatively small but too large to permit exhaustive testing of all possible combinations. In software testing, orthogonal arrays are effective in identifying errors related to faulty logic. In quality control, they are closely associated with the Taguchi methods, developed in the early 1950s and widely adopted by industries in Japan and later the United States. These methods have been instrumental in improving the quality of manufactured goods, reducing costs, and accelerating the time to market. Orthogonal arrays have also been applied in engineering, biotechnology, marketing,  advertising, and many other modern technological fields.

\section{Orthogonal Arrays as Fractional Factorial Experiments} \label{sec:ffe}

Orthogonal arrays were introduced for their use in fractional factorial experiments, and this remains their most important application in statistics. An $N \times k$ orthogonal array can be used to perform a fractional factorial experiment with $N$ runs and $k$ factors, where the number of levels for a factor is equal to the number of symbols for the corresponding column. Typically, the number of runs $N$ would be much smaller than the number of all possible level combinations for the $k$ factors, and if all runs of the orthogonal array are distinct, in which case it is called a {\em simple orthogonal array}, then its runs form a subset of those in a full factorial which consists of all possible level combinations for the factors. However, the definition of an orthogonal array does not require distinct runs, and we consider it to represent a fractional factorial as long as $N$ is smaller than the number of all possible level combinations. 

A well-known class of fractional factorials in which all factors have the same number of levels $s$ consists of the {\em regular fractional factorials} \citep[cf.][for $s=2$]{box19612}. Regular fractional factorials are a subclass of orthogonal arrays, and are closely related to {\em linear orthogonal arrays} \citep[cf.][Chapter 4]{hedayat2012orthogonal}. A fractional factorial is said to be regular if $N$ is a power of $s$, say $N = s^p$, there are $p$ factors so that every possible level combination appears once for these $p$ factors, and the columns for the other $k-p$ factors can be computed as explained below from the columns of the initial $p$ factors. The relationships that exist between different columns in a regular fractional factorial are called alias relationships. For example, consider the $OA(8,2^4,3)$ in Table~\ref{table:oa_8_4}. Any two or three columns of this orthogonal array are orthogonal in the sense that every possible level combination appears twice (for two columns) or once (for three columns). Selecting any three columns, the levels for the fourth column can then be obtained  by observing that $d_1 + d_2+d_3+d_4=0$ (modulo 2), where the four columns are denoted as $d_1,d_2,d_3,d_4$. Because of this relationship, the four-factor interaction is part of the defining relation of this fractional factorial. It also implies that, with all computations modulo 2, $d_1 = d_2+d_3+d_4$, $d_2 = d_1+d_3+d_4$, $d_3=d_1+d_2+d_4$ and $d_4=d_1+d_2+d_3$. This is to be interpreted as the main effect of the first factor being aliased with (i.e., indistinguishable from) the three-factor interaction of factors 2, 3, and 4; the main effect of the second factor being aliased with the three-factor interaction of factors 1, 3, and 4; and so on.
Every regular fractional factorial has a defining relation. The relationships between the factors are captured by the alias structure. For example, if we wish to construct a regular fractional factorial of 16 runs with 7 two-level factors, we can start with a full factorial of 16 runs and 4 factors. Denoting these factors by $A,B,C,D$, we can define the additional 3 factors using these four columns. Rather than using levels 0 and 1, for 2-level fractional factorials, it is common practice to use levels $-1$ and 1, which we will follow in this example. One possible choice for defining three remaining factors is $E= AB$, $F = BCD$, $G = AD$. As such, we obtain the defining relation $I = ABE = BCDF = ADG = ACDEF = BDEG = ABCFG = CEFG $, where $I$ represents a column of all ones. Here, $ABE$, $BCDF$, $ADG$, $ACDEF$, $BDEG$, $ABCFG$, and $CEFG$ are called {\em words}, and correspond to interactions in the defining relation that are equal to $I$. Multiplying each term in the defining relation by $A$, and using that $A^2=1$, we obtain that $A= BE = ABCDF = DG = CDEF = ABDEG = BCFG = ACEFG$. This is part of the alias structure, implying that if we try to estimate the main effect of factor $A$, we are really estimating $A+BE+ABCDF+DG + CDEF +ABDEG+ BCFG + ACEFG$. We may interpret this as the main effect of $A$ if the interactions in this sum can be assumed to be negligible. The complete alias structure of this fractional factorial is, \\ \vspace*{-0.38in}

{\tiny
\begin{eqnarray*}
A = BE = ABCDF = DG = CDEF = ABDEG = BCFG = ACEFG &
B = AE = CDF = ABDG = ABCDEF = DEG = ACFG = BCEFG\\
C = ABCE = BDF = ACDG = ADEF = BCDEG = ABFG = EFG &
D = ABDE = BCF = AG = ACEF = BEG = ABCDFG = CDEFG\\
E = AB = BCDEF = ADEG = ACDF = BDG = ABCEFG = CFG&
F = ABEF = BCD = ADFG = ACDE = BDEFG = ABCG = CEG\\
G = ABEG = BCDFG = AD = ACDEFG = BDE = ABCF = CEF&
AC = BCE = ABDF = CDG = DEF = ABCDEG = BFG = AEFG\\
AF = BEF = ABCD = DFG = CDE = ABDEFG = BCG = ACEG &
BC = ACE = DF = ABCDG = ABDEF = CDEG = AFG = BEFG\\
BD = ADE = CF = ABG = ABCEF = EG = ACDFG = BCFEFG &
BF = AEF = CD = ABDFG = ABCDE = DEFG = ACG = BCEG\\
BG = AEG = CDFG = AGD = ABCDEFG = DE = ACF = BCEF& 
CE = ABC = BDEF = ACDEG = ADF = BCDG = ABEFG = FG\\
CG = ABCEG = BDFG = ACD = ADEFG = BCDE = ABF = EF&
\end{eqnarray*}
}
%{\scriptsize
%\begin{eqnarray*}
%A = BCF = ABCDG = CEH =DFG=ABEFH = BDEHG =ACDEFGH & B = %ACF = CDG = ABCEH = ABDFG=EFH = ADEHG =BCDEFGH \\
%C = ABF = BDG = AEH = ACDFG=BCEFH = ABCDEHG =DEFGH & 
%D = ABCDF = BCG = ACDEH =  AFG=BDEFH = ABEHG =CEFGH\\
%E = ABCEF = BCDEG = ACH = ADEFG=BFH = ABDHG =CDFGH& 
%F = ABC = BCDGF = ACEFH=ADG=BEH = ABDEFHG =CDEGH \\
%G = ABCFG = BCD = ACEGH=ADF=BEFGH = ABDEH =CDEFH&
%H = ABCFF = BCDGF = ACE=ADFGF=BEF = ABDEG =CDEFG\\
%AB = CF = ACDG = BCEH=BDFG=AEFH = DEHG =ABCDEFGH&
%AC = CF = ABDG = EH=CDFG=ACBEFH = BCDEHG =ADEFGH\\
%AD = BCDF = ABCG = CDEH=FG=ABDEFH = BEHG =ACEFGH&
%AE = BCEF = ABCDEG = CH=DEFG=ABFH = BDHG =ACDFGH\\
%AF = BC = ABCDFG = CEFH=DG=ABEH = BDEFHG =ACDEGH &
%AG = BCFG = ABCD = CEHG=DF=ABEFGH = BDEH =ACDEFH\\
%AH = BCFH = ABCDGH = CE=DFGH=ABEF = BDEG =ACDEFG &\\
%\end{eqnarray*}}

Together with the defining relation, these 15 equations establish the relationships between the $2^7 = 128$ effects including the grand mean effect, main effects, two-factor interaction effects, and higher-order interaction effects. 
We refer to Chapters 4 and 5 of \cite{wu2011experiments}  and Chapter 9 of \cite{cheng2016theory} for in-depth discussions of regular fractional factorials. Although the relationships between the columns of a regular fractional factorial are simple, this simplicity comes at the cost of the run size having to be a power of the number of levels $s$. For example, for a two-level regular fractional factorial, the run size must be a power of 2, i.e. 4, 8, 16, 32, 64, and so on, thereby creating an increasingly large gap between any consecutive possible run sizes.   

The general class of fixed-level orthogonal arrays does not have such a constraint on the run size. For a two-level orthogonal array of strength two, for example, the run size can be any multiple of 4, that is, 4, 8, 12, 16, 20, and so on. Such arrays are known as non-regular fractional factorials if they are not regular. Their columns remain orthogonal (in the sense that all level combinations appear equally often as a row for any set of up to $t$ columns), but factors are now partially aliased (in the sense that there are no alias relationships as for regular factorials).  For instance, Table~\ref{table:oa_12_11} displays a two-level non-regular fractional factorial of 12 runs, 11 factors, and strength two. Any two columns of are orthogonal. But, for example, the levels for the factor corresponding to the third column cannot be computed from those of the factors corresponding to the first two columns (or any other two columns), because 00 for the first two factors can result in 0 or 1 for the third factor. Adding the levels for the first three factors modulo 2 results eight times in a sum of 0 and four times in a sum of 1, making the absolute correlation between the third factor and the interaction of the first two $\frac{1}{3}$. Since this absolute correlation is not 0 or 1, we call this partial aliasing. Similar observations hold for other sets of three columns.

Non-regular fractional factorials have a complex alias structure. Precisely because of this, non-regular fractional factorials were traditionally not favored in choosing fractions. However, with the advocacy of important work such as \cite{hamada1992analysis, lin1992projection,cheng1995some, deng1999generalized, hedayat2012orthogonal,dean2015handbook,chengtang2025}, the merits of non-regular fractional factorials are generally well recognized and they are widely used in the design and analysis of experiments, and beyond. 

\begin{table}[!h]
\begin{center}
\caption{An $OA(12,2^{11},2)$}
\begin{tabular}{ccccccccccc}
\multicolumn{11}{c}{}\\
0& 0& 0& 0& 0& 0& 0& 0& 0& 0& 0\\
0& 0& 0& 0& 0& 1& 1& 1& 1& 1& 1\\
0& 0& 1& 1& 1& 0& 0& 0& 1& 1& 1\\
0& 1& 0& 1& 1& 0& 1& 1& 0& 0& 1\\
0& 1& 1& 0& 1& 1& 0& 1& 0& 1& 0\\
0& 1& 1& 1& 0& 1& 1& 0& 1& 0& 0\\
1& 0& 0& 1& 1& 1& 1& 0& 0& 1& 0\\
1& 0& 1& 0& 1& 0& 1& 1& 1& 0& 0\\
1& 0& 1& 1& 0& 1& 0& 1& 0& 0& 1\\
1& 1& 0& 0& 1& 1& 0& 0& 1& 0& 1\\
1& 1& 0& 1& 0& 0& 0& 1& 1& 1& 0\\
1& 1& 1& 0& 0& 0& 1& 0& 0& 1& 1\\
\end{tabular}\label{table:oa_12_11} 
\end{center}
\end{table}
Definition \ref{def:OA} implies a key projection property of orthogonal arrays. Specifically, when an orthogonal array of strength $t$ is projected onto any subset of $t$ or fewer factors, the resulting subarray forms one or more replicates of the full factorial for those factors. 
This projection property has an important statistical implication: if there are at most $t$ active (i.e., important) factors, then an orthogonal array of strength $t$ allows for the estimation of all factorial effects of these active factors. Notably, this holds regardless of which factors are active.

\cite{deng1999generalized} introduced the concept of $J$-characteristics for studying two-level regular and non-regular fractional factorials. Let $D = [d_1,d_2,\ldots, d_k] = (d_{ij})$ denote an $N\times k$ two-level factorial with levels 1 and $-1$. For $1 \leq m \leq k$ and any $m$-subset $\mathcal{S} = \{d_{j_1},d_{j_2}, \ldots, d_{j_m}\}$ of the columns in $D$, \cite{deng1999generalized} defined
\begin{equation}\label{eq:J}
J_m(\mathcal{S}) = J_m(d_{j_1},d_{j_2}, \ldots, d_{j_m})= | \sum_{i=1}^N d_{ij_1} \cdots d_{ij_m}|.
\end{equation}
It can be shown that two-level orthogonal arrays of strength $t$ have $J_m(\mathcal{S}) = 0$ for all $m$-subsets $\mathcal{S}$ and $1 \le m \le t$. 
In addition, $D$ is a regular fractional factorial if and only if the $J$-characteristics of any subset is either 0 or $N$.  Thus, if the $J$-characteristic of some subset of a two-level fraction is strictly between 0 and $N$, then $D$ is a non-regular fractional factorial. 

An immediate question when using an orthogonal array as a fractional factorial is how to select the array (we will discuss this issue further in Section 7). In the pioneering work, \cite{box19612} introduced the concept of {\em resolution} to evaluate and compare two-level fractional factorials. In the context of regular two-level fractional factorials, they defined a fraction to be of resolution $R$ if no $c$-factor effect is aliased with any other effect containing less than $R-c$ factors. For example, a fraction of resolution III does not alias main effects with one another but allows main effects to be aliased with interactions of two or more factors, and a fraction of resolution IV does not alias main effects with each other or with two-factor interactions but does allow two-factor interactions to be aliased with each other. In two-level regular fractional factorials, its resolution is the length of the shortest word in the defining relation and is one more than the strength of the corresponding orthogonal array.  \cite{deng1999generalized} introduced {\em generalized resolution} to assess non-regular fractional factorials. Let $r$ be the smallest integer such that $\max_{|\mathcal{S}|=r} J_r(\mathcal{S}) >0$, where $J_r(\mathcal{S})$ is as defined in 
(\ref{eq:J}) and the maximization is over all $r$-subsets $S$ of $r$ distinct columns of $D$. \cite{deng1999generalized} defined the generalized resolution of design $D$ to be 
\begin{equation}\label{eq:R}
R(D) = r + [ 1 - \max_{|\mathcal{S}|=r} J_r(\mathcal{S})/N].   
\end{equation}
As the values of $J_r(\mathcal{S})$ are between $0$ and $N$ and the max in (\ref{eq:R}) is positive, we have $ r \leq R(D) < r+1$. A regular or nonregular fractional factorial of resolution $R$ is an orthogonal array of strength $\floor{R}$ -1 where $\floor{R}$ is the largest integer that does not exceed $R$. In view of this relationship, orthogonal arrays offer many attractive statistical properties when used as fractional factorials. For example, an orthogonal array of strength $t$ is universally optimal for any model consisting of factorial effects involving at most $\floor{t/2}$ factors  \citep{cheng1980orthogonal}. 

\section{Selected Applications of Orthogonal Arrays} \label{sec:applications}

Orthogonal arrays play a role in many applications. In Section~\ref{sec:ffe}, we have already explained the connection of orthogonal arrays to fractional factorials. Fractional factorials were introduced by \cite{finney1945fractional} and have found application in many fields, as already pointed out in Section~\ref{sec:definition}. Experiments using fractional factorials are especially prevalent in industry for product development and improvement and quality control. In this section, we highlight a few other applications with selected references for further reading.

\subsection{Application in Numerical  Integration} \label{subs:3.1}

%Due to their stratification properties and uniform projectivity, 
Orthogonal arrays are an important tool for numerical integration. This is due to the structure imposed by their definition. Considering the rows of an $s$-level orthogonal array of strength two as points in $k$-dimensional space, the array has the properties that (a) when projecting the points on any coordinate axis, the projections are uniformly distributed over $s$ equally spaced points; and (b) when projecting the points on any coordinate plane, the projections are uniformly distributed over $s \times s$ grids. In numerical integration, property (a) helps filter out the main effects, while property (b) is useful for filtering out two-factor interactions as well as main effects.  This will become clear as we define main effects and interaction effects below. 

Let $X = (X^1,X^2, \ldots, X^k)$ be uniformly distributed in $ [0,1]^k$ and $Y = f(X) \in \mathcal{R}$ where $f$ is a known, but expensive to compute, function. In numerical integration, a fundamental problem is how to select $N$ inputs $X_1,X_2, \ldots, X_N$ to approximate $\int f(X) dX$ accurately with a minimal sample size $N$. Often, $\int f(X) dX$ is estimated by $\bar{Y} = {\sum_{i=1}^N f(X_i)}/{N}$ where $f(X_i)$ is the value of the function evaluated at the input $X_i$. The main effect of the input variable $X^j$ is defined as $f_j(X^j) = E[f(X)|X^j] - E(Y)$ and the interaction effect between variable $X^i$ and variable $X^j$ is $f_{ij}(X^i,X^j) = E[f(X)|X^i,X^j] - E(Y) - f_i(X^i) - f_j(X^j)$, for $1 \leq i \neq j \leq k$. The simplest way to select $N$ inputs is random sampling and the corresponding variance  of $\bar{Y}$ is $N^{-1} \hbox{var}[f(X)]$. \cite{mckay1979comparison} introduced Latin hypercube sampling, which is  based on a Latin hypercube that is an array that, by definition, satisfies that each column is a permutation of $1, 2, \ldots, N$, where $N$ is the run size of the Latin hypercube.
\cite{mckay1979comparison} showed that Latin hypercube sampling achieves a smaller variance for $\bar{Y}$ than random sampling or stratified sampling when $f$ satisfies certain monotonicity conditions.  \cite{stein1987large} further derived that the variance of $\bar{Y}$ under Latin hypercube sampling is $N^{-1} \hbox{var}[f(X)]- N^{-1} \sum_{i=1}^k \hbox{var}[f_i(X^i)] + o(N^{-1})$ which is asymptotically smaller than the variance of $\bar{Y}$ under random sampling.  Rows of different Latin hypercubes for the same value of $N$ can have very different projections on two-dimensional coordinate planes. 
\cite{owen1992orthogonal} and \cite{tang1993orthogonal} independently took the idea in \cite{mckay1979comparison} further and introduced sampling methods based on orthogonal arrays to achieve further variance reduction for numerical integration. Their methods are based on {\em randomized orthogonal arrays} and {\em orthogonal array-based Latin hypercubes}, respectively. 
\cite{tang1993orthogonal} proved that the variance of $\bar{Y}$ under strength two orthogonal array-based Latin hypercube sampling is $N^{-1} \hbox{var}[f(X)]- N^{-1} \sum_{i=1}^k \hbox{var}[f_i(X^i)] - N^{-1}\sum_{i<j}^k \hbox{var}[f_{ij}]+ o(N^{-1})$.  Here we briefly explain the procedure for obtaining an orthogonal array-based Latin hypercube from an orthogonal array $D = OA(N,s^k,t)$.  
For each column of $D$, replace the $N/s$ positions with entry $u$, $u=0,1,...,s-1$, by a permutation of $uN/s + 1, uN/s + 2, \ldots, uN/s + N/s = (u+1)N/s$. The resulting array has the property that each column is a permutation of $1, 2, \ldots, N$ and thus a Latin hypercube \citep{mckay1979comparison}. Example~\ref{ex:oalhd} illustrates this procedure using an $OA(9,3^4,2)$. Properties of sampling by using such an orthogonal array-based Latin hypercube are discussed in \cite{tang1993orthogonal}, who refers to this procedure as {\em U sampling}. One key conclusion is that if the underlying function $f$ is additive, then U sampling gives a smaller variance of the integral approximation than the sampling in \cite{owen1992orthogonal}.
% , showing that the variance reduction property of Latin hypercube sampling extends to orthogonal array-based samples. 

\begin{exam}\label{ex:oalhd}
Consider $D$, an $OA(9,3^4,2)$, shown below. For each column of $D$, we replace level $0$ by a random permutation of $1,2,3$, level $1$ by a random permutation of $4,5,6$, and level $2$ by a random permutation of $7,8,9$. The orthogonal array-based Latin hypercube $L$ shown below is one of the arrays that we might obtain. 
$$ D = \left(\begin{array}{cccc}
0 & 0 & 0 & 0\\
0& 1& 1& 2\\
0& 2 & 2 & 1\\
1& 0 & 1 & 1\\
1& 1 & 2 & 0\\
1& 2 & 0 & 2\\
2& 0 & 2 & 2\\
2& 1 & 0 & 1\\
2& 2 & 1 & 0
\end{array}
\right), \  \ 
L = \left(\begin{array}{cccc}
1& 2 & 3 & 3\\
2& 6 & 4 & 9\\
3& 8 & 7 & 6\\
4& 3 & 6 & 4\\
5& 5 & 9 & 1\\
6& 7 & 1 & 7\\
7& 1 & 8 & 8\\
8& 4 & 2 & 5\\
9& 9 & 5 & 2
\end{array}
\right).
$$

\end{exam}

\subsection{Application in Computer Experiments} \label{subs:3.2}

With the exponential growth of computing power, researchers are increasingly using computer experiments to simulate real-world phenomena and complex systems through mathematical models. These models are solved using numerical methods such as computational fluid dynamics and finite element analysis to gain deeper insights into the systems being studied \citep{santner2003design,gramacy2020surrogates}.  The underlying mechanisms of these computer experiments are represented and executed through computer codes. To choose inputs to run computer codes, a widely used approach is to use space-filling designs which aim to spread out the design points evenly over the entire design space. Unlike traditional physical experiments, computer experiments often involve a larger number of input variables and require more runs. However, only a subset of these input variables is typically considered of primary importance. To identify the key input variables, space-filling designs with desirable low-dimensional projection properties are commonly used. These designs can be generated using orthogonal array-based Latin hypercubes, which were described in Subsection~\ref{subs:3.1}.
In addition, \cite{chen2022study} justified  orthogonal array-based designs under a broad class of space-filling criteria including commonly used distance-, orthogonality- and discrepancy-based measures. 

Orthogonal arrays are not only directly used in orthogonal array-based Latin hypercubes but also serve as a cornerstone for various constructions of space-filling designs. For instance, \cite{steinberg2006construction}, \cite{lin2009construction}, \cite{pang2009construction},  \cite{lin2016general},  \cite{sun2017general} and \cite{sun2017method} constructed another class of space-filling designs known as orthogonal Latin hypercubes, which are Latin hypercubes with the property that every two distinct columns have zero correlation. 
More specifically, for example,  let us consider how orthogonal arrays are used in \cite{lin2009construction}.  For any positive integer $u$, let $g_u$ be the $u \times 1$ vector
with $i$th element $i-(u+1)/2$, $1 \leq i \leq u$, and $\Gamma_u$ be
the set of the $u!$ vectors generated by permuting the elements of
$g_u$. Let $B = (b_{ij})$  be an $n \times p$  matrix with columns from
$\Gamma_n$. Suppose that an orthogonal array $OA( n^2, n^{2f},  2)$, say $A$, with
$n^2$ rows, $2f$ columns, $n$ symbols, and strength two is available. Denote the symbols in $A$ by $0, 1, 2, \ldots, n-1$. 
\cite{lin2009construction} proposed the following construction steps: 

\begin{enumerate}
\item[I.] For $1 \leq j \leq p$, obtain an $n^2 \times (2f)$
matrix $A_j$ by replacing the symbols $0, 1, 2, \ldots, n-1$ in
$A$ by $b_{1j}, b_{2j}, \ldots, b_{nj}$ respectively, and
then partition $A_j$ as $A_j$ = [$A_{j1}, \ldots, A_{jf}$], where
each of $A_{j1}, \ldots, A_{jf}$ has two columns.

\item[II.] For $1 \leq j \leq p$, obtain the $n^2 \times (2f)$
matrix $M_j$ = [$A_{j1}V, \ldots, A_{jf}V$] , where
\[
{
\renewcommand{\arraystretch}{0.6}
V = \left[
\begin{array}{rr}
1 & -n \\
n & 1 \\
\end{array}
\right]. }
\]

\item[III.] Finally, obtain the matrix $M$ = [$M_{1}, \ldots,
M_{p}$], of order $N \times q$, where $N=n^2$ and $q=2pf$.
\end{enumerate}

Step II of this construction applies a rotation to pairs of columns of the orthogonal array $A_j$. \cite{lin2009construction} showed that the resulting matrix $M$ is a Latin hypercube design in which the correlation between any two distinct columns depends only on correlations of columns of the matrix $B$. Thus, this construction can be used to obtain orthogonal Latin hypercubes of larger run sizes from those with smaller run sizes.  
  
In Section 8, we will discuss selected recent developments for orthogonal arrays. Some of these are motivated by applications in computer experiments. For example, sliced orthogonal arrays, introduced by \cite{qian2009sliced}, are a special class of orthogonal arrays with the property that runs can be partitioned into smaller orthogonal arrays, and these arrays are needed in computer experiments with both quantitative and qualitative inputs \citep{qian2008gaussian}.

\subsection{Application in Subsampling of Big Data}

With the advancement of technology, data generation continues to grow exponentially, potentially resulting in huge datasets. As a result, many fields, including statistical science, face unique challenges and unprecedented opportunities. One such challenge and opportunity is the development of subsampling methods for efficiently selecting subdata (i.e., a subset of a large dataset) with minimal loss of information. For example, \cite{wang2019information} proposed a novel approach, termed information-based optimal subdata selection (IBOSS), in the context of big data linear regression problems, and proposed a computationally efficient algorithm for approximating the optimal subdata via the IBOSS method. 
\cite{wang2021orthogonal} also considered a subsampling approach, but used orthogonal arrays to select optimal subdata. Consider the linear regression model with $p$ predictors or features,
$$ 
y = \beta_0 + \beta_1 x_1 + \beta_2 x_2 + \ldots + \beta_p x_p + \epsilon,
$$
\noindent 
where $y$ is the response and the random errors $\epsilon$ are independent and identically distributed with mean 0 and variance $\sigma^2$. The least squares estimate of $ \bbeta  = (\beta_0, \beta_1,\ldots, \beta_p)^T$ is $\hat{\bbeta}  = (\tilde{X}^T \tilde{X})^{-1} \tilde{X}^T \mathbf{y}$ where $\mathbf{y}$ is the response vector, $\tilde{X} = (1, X)$ and $X$ is the $N \times p$ matrix of feature values for the $N$ observations. Now take a subsample of size $n$ from the full dataset $(X,\mathbf{y})$ and let $(X_s,\mathbf{y}_s)$ denote this subsample. The least squares estimate based on the subsample is 
$$\hat{\bbeta}_s = (\tilde{X}_s^T \tilde{X}_s)^{-1} \tilde{X}_s^T \mathbf{y}_s,$$
\noindent where $\tilde{X}_s = (1, X_s)$. The covariance matrix of $ \hat{\bbeta}_s$ is $\sigma^2(\tilde{X}_s^T \tilde{X}_s)^{-1}$. 
Information-based subsampling approaches aim to find subdata that, in some way, minimize the variance of $\hat{\bbeta}_s$. Using an optimality function, say $\phi$, the optimal subdata minimizes $\phi ( (\tilde{X}_s^T \tilde{X}_s)^{-1} )$, i.e.,
$$
X_s^* = \argmin_{X_s \subset X} \phi ( (\tilde{X}_s^T \tilde{X}_s)^{-1} ).  
$$
Finding an exact solution for $N >> n$ is too expensive, so that algorithms seek a highly efficient solution for $X_s^*$.

Common choices for $\phi$ are the determinant and trace, which correspond to the criteria of $D$- and $A$-optimality, respectively, in optimal design of experiments.  The orthogonal subsampling proposed by  \cite{wang2021orthogonal} selects the subsample $X_s$ such that if all the covariates are scaled to [-1,1], $X_s$ mimics a two-level orthogonal array of strength two as closely as possible. The method is inspired by the optimality of orthogonal arrays for linear regression models. For example,  \cite{cheng1980orthogonal} showed that an orthogonal array of strength two with $s$ levels is universally optimal for a main-effects model. That is, such an array is optimal under a wide range of criteria that include $D$- and $A$-optimality, among all $s$-level factorial designs.  

A subsample $X_s$ for which the rows form exactly a two-level orthogonal array will generally not exist in the full dataset from which the sample is taken. \cite{wang2021orthogonal} introduced a discrepancy function that is to be minimized in order to sequentially select subdata that aims for the simultaneous attainment of two features. These two features are: (i) select points with extreme values of the features: selected points are located near the corners of the feature space and have a large distance from the center, and (ii) aim for orthogonality of columns corresponding to any two features. \cite{wang2021orthogonal} derived a lower bound for their discrepancy function and proposed an efficient algorithm to sequentially select points for inclusion in the subdata that minimize the discrepancy function. Interested readers are referred to their article for the details of this method and algorithm. 

\cite{zhang2024independence} and \cite{zhu2024group} extended the idea of the orthogonal subsampling to independence-encouraging subsampling for nonparametric additive models and group-orthogonal subsampling for hierarchical data based on linear mixed models, respectively. In both cases, orthogonal arrays serve as an essential tool for the proposed subsamplings. 

\section{Orthogonal Arrays and Error-Correcting Codes }

This section focuses on connections between orthogonal arrays and error-correcting codes. These two concepts have deep connections as first observed in \cite{bose1961some} for linear codes and linear orthogonal arrays. The latter are orthogonal arrays with a defining relation \citep[cf.][Section 11.5]{hedayat2012orthogonal}. An even deeper connection is based on the work by \cite{delsarte1973algebraic}. For further discussion, we begin by reviewing  fundamental concepts of an error-correcting code. Excellent references on this topic are \cite{macwilliams1977theory}, \cite{stinsontutorial} and \cite{hedayat2012orthogonal}. 
Error-correcting codes are used to detect and correct errors that occur during data transmission over noisy communication channels.  With a set of symbols $S$ of size $s$, called the {\em alphabet}, an {\em error-correcting code} is any collection $C$ of vectors from $S^k$, the set of all $s^k$ vectors of length $k$ based on the alphabet $S$. The vectors in $C$ are called the {\em codewords}. For codes it is not common that codewords are repeated (i.e., all vectors in $C$ are typically distinct), but since this is not a requirement for orthogonal arrays, we allow repetition of codewords for codes.

An important concept for a code $C$ is its {\em minimal distance}, which is defined as 
$$d = \min_{\stackrel{u,v \in C}{ u \neq v}} \hbox{dist}(u,v),$$
\noindent where $\hbox{dist}(u,v)$ is the number of positions where vectors $u$ and $v$ differ, referred to as the Hamming distance between $u$ and $v$. 
A code with minimal distance $d$ can correct $\lfloor (d-1)/2 \rfloor$ errors by associating a received signal with the word in the code that is closest to the signal in Hamming distance. If $C$ contains $N$ codewords, it is a code of length $k$, size $N$, and minimal distance $d$ over an alphabet of size $s$, denoted as a $(k,N,d)_s$ code. Table~\ref{table:code} provides a $(7,8,4)_2$ code.

When $S$ corresponds to a Galois field, we define a {\em linear code} $C$ of length $k$ as a code with distinct codewords that form a vector subspace of $S^k$. This definition implies that $C$ has size $N = s^n$ for some nonnegative integer $n$, $0\leq n \leq k$, where $k$ is now called the dimension of the code. A linear code may be characterized by an $n \times k$ {\em generator matrix} $G$. The rows of $G$ form a basis for the code, so that all codewords can be obtained by taking all possible linear combinations of the rows of $G$. For any linear code $C$, there is another linear code called its {\em dual}, and denoted by $C^\perp$. This consists of all vectors $v \in S^k$ such that $uv^T = 0$ for all $u \in C$. For a $(k,s^n,d)_s$ linear code $C$, its dual code $C^\perp$ is a $(k,s^{k-n},d^\perp)_s$ code, where $d^\perp$ is called the dual distance of $C$. Example~\ref{example:code} illustrates these concepts.  Common examples of linear codes include Hamming codes, Bose-Chaudhuri-Hocquenghem (BCH) codes, Reed-Solomon codes, cyclic codes, Golay codes, and Reed-Muller codes. \cite{hedayat2012orthogonal} devoted a chapter discussing the construction of orthogonal arrays using these codes. 

Nonlinear codes have been investigated much less than linear codes, but there are families of nonlinear codes that tend to have better encoding properties than linear codes of the same size. \cite{nordstrom1967optimum} provided the first nonlinear code now known as the Nordstrom-Robinson code. It is a $(16,256,6)_2$ code with the property that it has dual distance 6 and offers the advantage over linear codes in that any binary linear code of length 16 with minimal distance 6 can contain at most 128 codewords. For development of families of nonlinear codes that generalize the Nordstrom-Robinson code we refer to \cite{macwilliams1977theory}.

\begin{exam}\label{example:code}
Table \ref{table:code} lists a $(7,8,4)_2$ code. Each row corresponds to a codeword. The generator matrix of this code is
$$G = \left[ \begin{array} {ccccccc}
1 & 1 & 1 & 0 & 1 & 0 & 0  \\
0 & 1 & 1 & 1 & 0 & 1 & 0\\
0 & 0 & 1 & 1 & 1 & 0 & 1 \\
\end{array}
\right]. $$
Its dual code is a $(7,16,3)_2$ code which has the generator matrix,
$$ \left[ \begin{array} {ccccccc}
1 & 0 & 1 & 1 & 0 & 0 & 0 \\
0 & 1 & 0 & 1 & 1 & 0 & 0 \\
0 & 0 & 1 & 0 & 1 & 1 & 0 \\
0 & 0 & 0 & 1 & 0 & 1 & 1 
\end{array}
\right]. $$

\begin{table}[!h]
\begin{center}
\caption{A $(7,8,4)_2$ code}
\begin{tabular}{ccccccc}
\multicolumn{7}{c}{}\\
0 & 0 & 0 & 0 & 0 & 0 & 0 \\
1 & 1 & 1 & 0 & 1 & 0 & 0  \\
0 & 1 & 1 & 1 & 0 & 1 & 0\\
0 & 0 & 1 & 1 & 1 & 0 & 1 \\
1 & 0 & 0 & 1 & 1 & 1 & 0\\
0 & 1 & 0 & 0 & 1 & 1 & 1\\
1 & 0 & 1 & 0 & 0 & 1 & 1\\
1 & 1 & 0 & 1 & 0 & 0 & 1\\
\end{tabular}\label{table:code} 
\end{center}
\end{table}
\end{exam}

\subsection{Basic Relationship Between Orthogonal Arrays and Codes}
We define an orthogonal array $OA(N,s^k,t)$ based on the Galois field $S = GF(s)$ to be linear if its runs are distinct and form a vector space over $S^k$. This is identical to the definition for a linear code. Consequently, a linear orthogonal array $OA(N,s^k,t)$ is a linear code $(k,N,d)_s$ for some $d$, and vice versa, a linear code $(k,N,d)_s$ is a linear orthogonal array $OA(N,s^k,t)$ for some strength $t$. It turns out that this relationship connects the strength of a linear orthogonal array to the dual distance of a code. The relationship is formulated precisely in Theorem~\ref{thm:code} below, which was first stated by \cite{bose1961some} \citep[see also][for a statement and proof]{hedayat2012orthogonal}. 
Through this relationship it is possible to obtain linear orthogonal arrays from linear codes, and vice versa. Existence results can also be translated from orthogonal arrays to codes and vice versa. 
\begin{thm}\label{thm:code}
If $C$ is a $(k,N,d)_s$  linear code over $S = GF(s)$ with dual distance $d^\perp$, then the codewords of $C$ form the rows of an $OA(N,s^k,d^\perp-1)$ with entries from $GF(s)$. Conversely, the rows of a linear $OA(N,s^k,t)$ over $GF(s)$ form a $(k,N,d)_s$ linear code over $GF(s)$ with dual distance $d^\perp \geq t+1$. If the orthogonal array has strength $t$ but not $t+1$, then $d^\perp = t+1$.
\end{thm}

% \begin{defn}
% Let $s$ be a prime power. An orthogonal array $OA(N,s^k,t)$ with the levels from $GF(s)$is said to be {\em linear} if its runs are distinct and if its $N$ runs form a vector space over $GF(s)$. 
% \end{defn}

It should be noted that a linear orthogonal array corresponds to a regular fractional factorial and that the words in the defining relation correspond to the rows in the dual of this linear orthogonal array.

\subsection{Rao's Bound and the Linear Programming Bound}
Two very basic and related problems associated with the existence of orthogonal arrays are the following:
\begin{enumerate}
    \item For given values of $k$, $s$ and $t$, what is the smallest number of runs $N$ for which an $OA(N,s^k,t)$ exists?
    \item For given values of $N$, $s$ and $t$, what is the largest number of factors $k$ for which an $OA(N,s^k,t)$ exists?
\end{enumerate}
It can be seen that a complete answer to the second question implies a complete answer to the first question \citep[cf.][Chapter 2]{hedayat1992asymmetrical}.

Unfortunately, exact answers to these questions are often unknown. \cite{rao1947factorial} provides a lower bound for $N$ in terms of $k$, $s$ and $t$ that applies to any orthogonal array. The bound is now also known as Rao's bound. Implicitly, Rao's bound also provides an upper bound for the value of $k$ for given $N$, $s$ and $t$. Rao's bound states that for an $OA(N,s^k,t)$ it must hold that
\begin{eqnarray*}
N \ge \sum_{i=0}^u {k \choose i} (s-1)^i, \text{~~if }  t = 2u, \text{~~and}\\
N \ge \sum_{i=0}^u {k \choose i} (s-1)^i + {k-1 \choose u}(s-1)^{u+1}, \text{~~if } t = 2u+1,
\end{eqnarray*}
for $u \ge 0$. This result can be understood by counting degrees of freedom for main effects and interaction effects that can be estimated orthogonally when the orthogonal array is used in a fractional factorial experiment.

While improvements on Rao's bound were found for special cases, it wasn't until the seminal work by \cite{delsarte1973algebraic} that another general bound was established, which is known as the linear programming bound. Unlike Rao's bound, the linear programming bound provides a bound through computation (by linear programming) and does not provide an explicit lower bound for $N$. Theorem~\ref{thm:bound} presents the linear programming bound. 

\begin{thm}\label{thm:bound}
    Let $N_{LP}(k,d^\perp)$ be the solution to the following linear programming problem: find real numbers $A_0, A_1, \ldots,A_k$ to minimize
    $$A_0+A_1 + \cdots +A_k,$$
\noindent subject to the constraints
$$\begin{array}{lll}
A_0 & \geq & 1, \ \ A_i \geq 0,\ \  1 \leq i \leq k\\
B_0 & = & 1, \ \ B_i \geq 0, \ \ 1 \leq i \leq k \\
B_1 & = & \dots \ \ = \ \ B_t \ \ = 0,
\end{array}$$
\noindent where 
$B_i = \sum_{j=0}^k A_j P_i(j), 0 \leq i \leq k$, the $P_i(j)$ are the Krawtchouck polynomials 
$P_i(j) = \sum_{r=0}^i (-1)^r (s-1)^{i-r}{j \choose r}
{k-j \choose i-r}$, and $ t= d^\perp - 1 $. Then, in an $OA(N,s^k,t)$, it holds that 
$$ N \geq N_{LP}(k,d^\perp).$$
\end{thm}

\cite{delsarte1973algebraic} demonstrated that the Rao's bound follows as a consequence of Theorem~\ref{thm:bound}, establishing that the linear programming bound is always at least as strong as the Rao's bound. In many cases, the linear programming bound is significantly stronger. \cite{hedayat2012orthogonal} provided a comparative table highlighting the differences between the Rao's and linear programming bounds for binary orthogonal arrays of strength 4 with $k$ factors. Building on the work of \cite{delsarte1973algebraic}, \cite{sloane1996linear} extended these results and formulated the linear programming bound for mixed-level orthogonal arrays. Notably, Table 9.7 in \cite{hedayat2012orthogonal} presents the linear programming bounds for $OA(N, 2^{k_1}3^{k_2},t)$'s, demonstrating significant improvements over the Rao's bound extended to mixed orthogonal arrays.

\section{Connections Between Orthogonal Arrays and Other Combinatorial Structures}

In this section, we will briefly explore connections between orthogonal arrays and other combinatorial structures, such as mutually orthogonal Latin squares, Hadamard matrices, incomplete block designs and difference schemes. Interested readers are referred to Chapters 6, 7 and 8 in \cite{hedayat2012orthogonal}.

\subsection{Mutually Orthogonal Latin Squares}

Interest by mathematicians in Latin squares and mutually orthogonal Latin squares dates back to the early 1700s. Their use in statistics was discovered in the 1930s largely due to the influential work of R.A. Fisher. In particular, Latin squares and mutually orthogonal Latin squares can play a role in experiments for comparing different treatments in the presence of multiple blocking variables. A {\em Latin square} of order $s$ is an $s \times s$ array  with entries of a set of $S$ of size $s$ such that each element of $S$ appears once in every row and column. For example, the following three arrays are Latin squares of order 4: 
 
$$
\begin{array}{cccc c  cccc c cccc}
0 & 2 & 3 & 1 & \qquad \qquad \qquad & 0 & 2 & 3 & 1 & \qquad \qquad \qquad & 0 & 2 & 3 & 1 \\
3 & 1 & 0 & 2 & \qquad \qquad \qquad & 1 & 3 & 2 & 0 & \qquad \qquad \qquad & 2 & 0 & 1 & 3 \\
1 & 3 & 2 & 0 & \qquad \qquad \qquad & 2 & 0 & 1 & 3 & \qquad \qquad \qquad & 3 & 1 & 0 & 2 \\
2 & 0 & 1 & 3 & \qquad \qquad \qquad & 3 & 1 & 0 & 2 & \qquad \qquad \qquad & 1 & 3 & 2 & 0 \\
\end{array}
$$
 
Two Latin squares of order $s$ are said to be {\em orthogonal} to each other if, when superimposed on each other, each of the $s^2$ pairs $(i,j)$ appears in exactly one cell, for $1 \leq i, j \leq s$. It can be verified that any two of the above three Latin squares of order 4 are orthogonal to each other. A set of {\em mutually orthogonal Latin squares} is a collection of Latin squares of order $s$ in which any pair is orthogonal. The existence and construction of mutually orthogonal Latin squares are discussed in \cite{hedayat2012orthogonal}.
The primary connection between mutually orthogonal Latin squares and orthogonal arrays is summarized in
Theorem \ref{thm:MOLS}. The proof can be found in \cite{hedayat2012orthogonal} and \cite{cheng2016theory}. 

\begin{thm}\label{thm:MOLS}
A set of $k$ mutually orthogonal $s \times s$ Latin squares is equivalent to an $OA(s^2,s^{k+2},2)$. 
\end{thm}

For example, the three mutually orthogonal Latin squares of order 4 are equivalent to an $OA(16,4^5,2)$. It is known that an upper bound on the number of mutually orthogonal Latin squares of order $s$ is $s-1$ and this upper bound can be achieved when $s$ is a prime or prime power.  
Even with today's computing power, the problem of obtaining the maximum possible number of mutually orthogonal Latin squares for other values of $s$ remains, with few exceptions, a challenging undertaking. The study of this problem was pioneered by \cite{bose1960further} and \cite{wilson1974concerning}, and pursued by many others. Interested readers are referred to the brief survey provided by \cite{colbourn2001mutually}. We note that although the connection between mutually orthogonal Latin squares and orthogonal arrays is fascinating, the use of mutually orthogonal Latin squares for constructing new orthogonal arrays may be limited.  \cite{hedayat2012orthogonal} described a number of interesting related research problems to be addressed.

\subsection{Hadamard Matrices} \label{subs:hadamard}

The concept of a Hadamard matrix originated in 1893
from the work of the French mathematician Jacques Salomon Hadamard. The study of Hadamard matrices expanded significantly in the 20th century, particularly with their applications in coding theory, signal processing, and  design of experiments.  Formally, a Hadamard matrix is a square matrix whose entries are either 1 or $-1$ and whose rows (and, hence, columns) are mutually orthogonal. Mathematically, a Hadamard matrix $H$ satisfies, 
$$H H^T = N I_N$$
\noindent where $N$ is the order of the matrix and $I_N$ is the $N \times N$ identity matrix.  Theorem~\ref{thm:HadaOA} establishes a connection between Hadamard matrices and orthogonal arrays. Multiplication of entire rows or columns of a Hadamard matrix by $-1$ will again result in a Hadamard matrix, so that for every order $N$ for which a Hadamard matrix exists, there is one with all entries in the first row or column equal to 1.  

\begin{thm}\label{thm:HadaOA}
 Suppose $H$ is a Hadamard matrix of order $N >2$ such that  all entries in its first column are 1. If the first column of 
$H$ is removed, the resulting matrix is an  $OA(N,2^{N-1},2)$. Conversely, appending a column of all 
1's to an $OA(N,2^{N-1},2)$ produces a Hadamard matrix of order $N$.
\end{thm}
The proof of Theorem~\ref{thm:HadaOA} can be found in \cite{hedayat2012orthogonal} and \cite{cheng2016theory}. Since the existence of an $OA(N,2^{N-1},2)$ is equivalent to the existence of an $OA(2N,2^N,3)$, the existence of the latter orthogonal array is also equivalent to the existence of a Hadamard matrix of order $N$ \citep[cf.][]{hedayat1978hadamard, hedayat2012orthogonal}. If there exists a Hadamard matrix of order $N>2$, then $N$ must be a multiple of 4.  According to the Hadamard conjecture, Hadamard matrices exist for all orders that are multiples of 4, but their existence has not been proven for all such orders. The smallest multiple of 4 for which no Hadamard matrix has been found is 668, despite extensive computational searches. Given the equivalence between Hadamard matrices and orthogonal arrays, it is clear that two-level orthogonal arrays can be constructed by selecting specific columns from Hadamard matrices.
Orthogonal arrays obtained from Hadamard matrices, also called Hadamard designs, are not regular when the order $N$ of the Hadamard matrix is not a power of 2; when $N$ is a power of 2, they may be either regular or non-regular. This flexibility allows Hadamard designs to accommodate a wider range of run sizes compared to regular designs. In their seminal work, \cite{plackett1946design} introduced the use of Hadamard designs in factorial experiments. The Hadamard designs described in their paper are now commonly known as {\em Plackett-Burman designs}.

Several well-established methods have been introduced to construct Hadamard matrices. These include the Sylvester construction, Paley construction, Williamson construction, conference matrices, and algebraic approaches involving group theory or combinatorial designs
\citep{sylvester1867lx,paley1933orthogonal, williamson1944hadamard, hall1998combinatorial, georgiou2003hadamard,kline2019geometric}.  For small orders, Hadamard matrices can be found through exhaustive search or optimization algorithms. 
Several researchers have conducted theoretical investigations into identifying suitable Hadamard matrices for constructing two-level non-regular designs \citep{shi2018designs, chen2023nonregular}. In addition, Hadamard matrices have been used to construct mixed orthogonal arrays through clever use of properties of Hadamard matrices
\citep{dey1977note,chacko1979orthogonal,chacko1981some,agrawal1982note,cheng1989shorter,wang1990constructions}. See Section 4.3 of \cite{dey2009fractional} for detailed construction methods and results. 

\subsection{Incomplete Block Designs }

Blocking is one of the three fundamental principles in the design of experiments. It is an effective strategy for explaining response variability by controlling for known nuisance factors, thereby increasing the precision of treatment effect estimation \citep{wu2011experiments}. The simplest and most frequently used block designs are randomized complete block designs. However, blocking should be based on differences in experimental units, and it may not always be feasible to include all treatments within a single block. {\em Incomplete block designs} may then be a sensible choice. Yates introduced   {\em balanced incomplete block designs}, for which $v$ treatments are arranged in $b$ blocks of $k$
experimental units, $k<v$, each treatment occurring in $r$ blocks, and any two treatments occurring together in $\lambda$ blocks. Each balanced incomplete block design has therefore the parameters: $v$ (the number of treatments); $b$ (the number of blocks); $r$ (the number of blocks in which each treatment appears); $k$ (the number of treatments in each block); $\lambda$ (the number of blocks in which any pair of treatments appears together). 
Necessary conditions for the existence of a balanced incomplete block design are that the parameters satisfy $bk = vr$, $\lambda(v-1) = r(k-1)$ and $b \geq v$. If $b=v$, then $r=k$ and the design is called a symmetric balanced incomplete block design \citep[cf.][]{lander1983symmetric}.

The connection between orthogonal arrays and incomplete block designs is partly due to both being connected to Hadamard matrices \citep{hedayat1978hadamard}. One connection between orthogonal arrays and incomplete block designs is stated in the following theorem.

\begin{thm} \label{thm:OABIB}
The existence of an orthogonal array $OA(N,2^{N-1},2)$ implies the existence of a symmetric balanced incomplete block design with $v=b=N-1$, $r=k=N/2-1$, and $\lambda=N/4-1$.
\end{thm}

To see the validity of Theorem~\ref{thm:OABIB}, we represent an incomplete block design through its incidence matrix, say $M=(m_{ij})$. This is a $v \times b$ matrix with entries 0 and 1, and with $m_{ij}=1$ if and only if the $i$th treatment appears in the $j$th block. Starting from the orthogonal array, we obtain a Hadamard matrix of order $N$ (see Theorem~\ref{thm:HadaOA}), and convert it to one in which all entries in both the first column and row are 1. The matrix that remains after deleting the first column and row is the incidence matrix for a symmetric balanced incomplete block design with the parameters as in Theorem~\ref{thm:OABIB}. The converse also holds, so that the existence of this symmetric balanced incomplete block design is equivalent to the existence of an orthogonal array $OA(N,2^{N-1},2)$. 

There are multiple other block designs that are related to the incomplete block design in Theorem~\ref{thm:OABIB}. For example, by replacing each block by its complement in the treatment set we obtain a symmetric balanced incomplete block design with $v=b=N-1$, $r=k=N/2$, and $\lambda=N/4$. As another example, starting from the symmetric block design in Theorem~\ref{thm:OABIB}, upon deleting an entire block and deleting the treatments in that block also from all of the other blocks, we obtain the so-called residual design, which is a balanced incomplete block design with $v=N/2$, $b=N-2$, $r=N/2-1$, $k=N/4$, and $\lambda=N/4-1$. Additional connections can be found in \cite{hedayat1978hadamard} and \cite{raghavarao1988constructions}.

\subsection{Difference Schemes}

Difference schemes were first defined by \cite{bose1952orthogonal}. We review their concept and their connection with orthogonal arrays. More importantly, they are a simple yet powerful tool for constructing orthogonal arrays as we will see in Section~\ref{sec:construction}. 

An $r\times c$ array with entries from a finite Abelian group containing $s$ entries is called a {\em difference scheme} if each vector difference between any two distinct columns of
the array contains every element from the group 
equally often \citep{bose1952orthogonal}. Such an array is denoted by  $D(r, c, s)$.   Table~\ref{table:differencescheme} displays a $D(9,9,3)$.  \cite{hedayat2012orthogonal} discussed the existence, construction, properties, and generalizations of difference schemes. 
Statistical Analysis Software (SAS) provides a library of difference schemes with a large number of different values for $r$ and $c$ and $3 \le s \le 22$. The library can be accessed at the webpage \cite{sas}. In this definition, difference schemes are of strength two. \cite{hedayat1996difference} introduced difference schemes of strength $t$ for the integer $t \geq 2$. 

Every orthogonal array $OA(N,s^k,2)$ based on an Abelian group is a difference scheme $D(N,k,s)$, but typically not a very interesting one. When looking for a difference scheme $D(r,c,s)$ one is typically interested in obtaining an array with the largest possible value of $c$ for given values of $r$ and $s$, and the additional structure required by an orthogonal array will fail to provide an array with the maximum possible value for $c$. 

For $s=2$, using the multiplicative Abelian group consisting of 1 and $-1$, a Hadamard matrix of order $N$ is a difference scheme $D(N,N,2)$ and, conversely, such a difference scheme can be converted to a Hadamard matrix of order $N$. Thus, through Hadamard matrices, this provides a connection between orthogonal arrays and difference schemes. While this is a simple and interesting connection, what is perhaps more interesting is that difference schemes can be used to construct larger arrays that are orthogonal arrays. For example, as we will see in Subsection~\ref{subs:difference}, the existence of a difference scheme $D(\lambda s, k, s)$ implies that of an orthogonal array $OA(\lambda s^2, s^{k+1},2)$
and a mixed orthogonal array $OA(\lambda s^2, (\lambda s)^1 s^{k},2)$. These orthogonal arrays can therefore be presented via the smaller difference schemes, and exploring the construction of the difference schemes, whether through combinatorial or computational methods, can be easier than direct constructions of orthogonal arrays. On the flip side, not every orthogonal array can be constructed through a difference scheme.

If $s$ is a prime or prime power, then $D(s,s,s)$ can be obtained based on the Galois field $GF(s)$, so that can we obtain an $OA(s^2,s^{s+1},2)$, which provides equality in the Rao's bound. \cite{hedayat2012orthogonal} studied an extension to difference schemes of strength $t$, $t \ge 3$, established the existence of such arrays, and showed how they could be used for the construction of orthogonal arrays of strength $t$. Difference schemes have also been used for the construction of mixed orthogonal arrays \citep[cf.][]{wang1991approach}.

\begin{table}[!h]
\begin{center}
\caption{A difference scheme $D(9,9,3)$}
\begin{tabular}{ccccccccc}
\multicolumn{9}{c}{}\\
0 & 0 & 0 & 0 & 0 & 0 & 0 & 0 & 0 \\
0 & 1 & 2 & 0 & 1 & 2 & 0 & 1 & 2 \\
0 & 2 & 1 & 0 & 2 & 1 & 0 & 2 & 1\\
0 & 0 & 0 & 2 & 2 & 2 & 1 & 1 & 1\\
0 & 1 & 2 & 2 & 0 & 1 & 1 & 2 & 0\\
0 & 2 & 1 & 2 & 1 & 0 & 1 & 0 & 2\\
0 & 0 & 0 & 1 & 1 & 1 & 2 & 2 & 2 \\
0 & 1 & 2 & 1 & 2 & 0 & 2 & 0 & 1 \\
0 & 2 & 1 & 1 & 0 & 2 & 2 & 1 & 0
\end{tabular}\label{table:differencescheme} 
\end{center}
\end{table}

\section{Selected Construction Methods for Orthogonal Arrays} \label{sec:construction}

This section provides a concise review of selected construction methods for orthogonal arrays. The literature on methods for constructing orthogonal arrays is extensive, encompassing a wide range of approaches. These include direct construction methods and methods based on other mathematical structures such as Hadamard matrices, finite fields, Latin squares, difference sets, cyclic groups, and projective planes. Recursive methods, such as those utilizing Kronecker products or Kronecker sums, are also widely used. In addition, intelligent computer-based searches have become an important tool for generating orthogonal arrays tailored to specific requirements. Here, we review several general and widely recognized methods.

\subsection{Constructions Based on Difference Schemes} \label{subs:difference}

Theorem~\ref{thm:diff} indicates that a difference scheme can be converted into an orthogonal array straightforwardly. 

\begin{thm}\label{thm:diff}
If $D$ is a difference scheme $D(r,c,s)$ based on an Abelian group $\mathcal{A} = \{\sigma_0,\sigma_1, \ldots, \sigma_{s-1}\}$ with binary operation $+$, then
$$A = \left [ 
\begin{array}{c}
D_0\\
D_1\\
\vdots\\
D_{s-1}
\end{array}
\right ]
$$ 
\noindent is an $OA(rs,s^c,2)$, where $D_i$ is obtained from $D$ by adding $\sigma_i$ to each of its entries.
\end{thm}
The orthogonal array in Theorem~\ref{thm:diff} can be extended by at least one column. For this we use the Kronecker product of an $n_1 \times m_1$ matrix $A=(a_{ij})$ and an  $n_2 \times m_2$ matrix $B$ that are based on the same group with binary operation $+$. Their Kronecker product is denoted by $A \otimes B$, and is defined as the $(n_1n_2) \times (m_1m_2)$ block matrix with the block in location $(i,j)$ equal to $a_{ij}+B$, $1 \le i \le n_1, ~1\le j \le m_1$. If $\sigma_0$ denotes the identity element of the group, then the column that can be added to the array in Theorem~\ref{thm:diff} is $\sigma_0 1_{r/s} \otimes (\sigma_0,...,\sigma_{s-1})^T$, where $\sigma_0 1_{r/s}$ denotes the $r/s \times 1$ vector with every entry equal to $\sigma_0$ and $(\sigma_0,...,\sigma_{s-1})^T$ is the $s \times 1$ vector that contains every element of $\mathcal{A}$. Thus, a difference scheme $D(r,c,s)$ implies the existence of an orthogonal array $OA(rs,s^{c+1},2)$. This explains the claim in Subsection~\ref{subs:difference} that a difference scheme $D(s,s,s)$ results in an $OA(s^2,s^{s+1},2)$. 

Kronecker products are a workhorse in constructing orthogonal arrays from difference schemes. Another use is outlined in Theorem~\ref{thm:diffOA}. This theorem states that a difference scheme, combined with an existing orthogonal array, can be used to generate an orthogonal array with a larger run size and more columns. 
% The construction uses an operator called Kronecker product. Let $A=(a_{ij})$ be an $n_1 \times m_1$ matrix and $B$ be an $n_2 \times m_2$ matrix, where both matrices have entries from the Galois field $GF(s)$. Throughout, we let $GF(s) = \{ \alpha_0, \alpha_1, \ldots, \alpha_{s-1}\}$ with $\alpha_0 = 0$ and $\alpha_1 = 1$. The {\em Kronecker product} of $A$ and $B$ is an $(n_1n_2) \times (m_1m_2)$ matrix given by
% \begin{equation}\label{eq:ks}
% A \otimes B= [{a_{ij}*{B}}]_{1\leq i\leq n_1, 1\leq j\leq m_1},
% \end{equation}
% \noindent where ${a_{ij}*{B}}=(a_{ij}*b_{pq})$ is an $n_2 \times m_2$  matrix with $*$ representing the addition in a field. 

\begin{thm}\label{thm:diffOA}
If $D$ is a difference scheme $D(r,c,s)$ and $B$ is an $OA(N,s^k,2)$, both based on the same Abelian group, then the array
$$A = B \otimes D $$
\noindent is an orthogonal array $OA(Nr, s^{kc},2)$.
\end{thm}

Additional methods for using difference schemes in the construction of orthogonal arrays can be found in Chapter 6 of \cite{hedayat2012orthogonal}. The concept was generalized by \cite{seiden1954problem}, who introduced difference schemes of strength $t$. Subsequently, \cite{hedayat1996difference} examined their existence and construction, providing methods to construct orthogonal arrays of strength $t$ from these schemes for $t >2$.

\subsection{Constructions Based on Hadamard Matrices}
In Theorem~\ref{thm:HadaOA} we already saw how a Hadamard matrix of order $N$ can be used to construct an orthogonal array $OA(N,2^{N-1},2)$. In addition, if $H$ is a Hadamard matrix, then the array
\[
\left[ \begin{array}{c}
     H  \\
     -H 
\end{array} \right]
\]
is an orthogonal array $OA(2N,2^N,3)$. Both arrays give equality in the Rao's bound. Thus, the construction of these orthogonal arrays boils down to the construction of Hadamard matrices. For the same order, different Hadamard matrices can result in orthogonal arrays with different properties. For example, if $N$ is a power of 2, we could obtain a regular or non-regular fractional factorial depending on the Hadamard matrix that we use. 

The simplest construction method for Hadamard matrices of order $N=2^m$, $m \ge 2$, is the Sylvester method, which is based on repeated use of the Kronecker product (with the binary operation being multiplication), as defined in Subsection~\ref{subs:difference}. This result is formally stated in Theorem~\ref{thm:saturatedOA}.
\begin{thm}\label{thm:saturatedOA}
Let $H$ be an array obtained by the $(m-1)$-fold Kronecker product, $m \ge 2$,
$$ H = \left[
\begin{array}{rr}
1 & -1 \\
1 & 1 
\end{array}
\right] 
\otimes \cdots \otimes 
 \left[
\begin{array}{rr}
1 & -1 \\
1 & 1 
\end{array}
\right]. 
$$
Then $H$ is a Hadamard matrix of order $2^m$.
\end{thm}

More generally, for Hadamard matrices $H_1$ and $H_2$ of order $N_1$ and $N_2$, their Kronecker product $H_1 \otimes H_2$ is a Hadamard matrix of order $N_1N_2$.

Based on Galois fields, \cite{paley1933orthogonal} developed two methods of construction for infinite families of Hadamard matrices. The first Paley construction works when $q = N-1$ is an odd prime power, where $N$, a multiple of 4, is the order of the Hadamard matrix to be constructed. Denote the elements of a Galois field $GF(q)$ by $\alpha_1 = 0, \alpha_2, \ldots, \alpha_q$, and define a function $\chi: GF(q) \rightarrow \{0,1,-1\}$ as 
$$
\chi (\beta) =  \begin{cases}
1, & \hbox{if } \beta = x^2 \ \hbox{for some } x \in GF(q);\\
0, & \hbox{if } \beta = 0; \\
-1, & \hbox{otherwise.}
\end{cases}$$
\noindent Let $A=(a_{ij})$ be the $q \times q$ matrix with $a_{ij} = \chi(\alpha_i - \alpha_j)$ for $i,j = 1, 2, \ldots, q$, and define
\begin{equation}\label{eq:paley1}
H =  
\begin{bmatrix}
    1 \ & -1_q^T\\
    1_q  \ & A + I_q\\
\end{bmatrix}
\end{equation}
\noindent where $1_q$ is a column of $q$ ones and $I_q$ is the $q \times q$ identity matrix.  The matrix $H$ in (\ref{eq:paley1}) is a Hadamard matrix of order $N$ \citep[cf.][]{hedayat2012orthogonal,cheng2016theory}. The $OA(N,2^{N-1},2)$ constructed from $H$ in (\ref{eq:paley1}) is also known as the {\em Paley design} of order $N$.  

The second Paley construction is for Hadamard matrices of order $N = 2q +2$, where $q$ is an odd prime power and $q-1$ is a multiple of 4. Using the matrix $A$ as defined for the first Paley construction, define
\begin{equation}\label{eq:paley2}
H= 
\begin{bmatrix}
1 & 1_q^T & -1 & 1_q^T\\
1_q & A +I_q & 1_q & A - I_q\\
-1 & 1_q^T & -1 & -1_q^T\\
1_q & A - 1_q & -1_q & -A - I_q
\end{bmatrix}.
\end{equation}
Then $H$ in (\ref{eq:paley2}) is a Hadamard matrix of order $N$.  
Multiplying the $(q+2)$nd row of $H$ in (\ref{eq:paley2}) by $-1$ gives a Hadamard matrix with first column equal to 1, which can be used to construct an orthogonal array $OA(N,2^{N-1},2)$. \cite{chen2023nonregular} conducted a comprehensive study of orthogonal arrays constructed from Hadamard matrices obtained by the Paley constructions in terms of generalized resolution, projectivity, and hidden projection property.

\subsection{The Rao-Hamming Construction}

Multiple methods of construction for orthogonal arrays make use of Galois fields or finite geometries. One of the earlier and simpler methods to describe was introduced independently by \cite{rao1947factorial,rao1949class} and \cite{hamming1950error}, and the corresponding orthogonal arrays were named Rao-Hamming arrays by \cite{hedayat2012orthogonal}. There are actually multiple methods of construction for the Rao-Hamming arrays \citep[see][for three methods]{hedayat2012orthogonal}, and we present one of these methods here. Rao-Hamming arrays are orthogonal arrays $OA(s^n, s^k,2)$ where $s$ is a prime power, $k=(s^n-1)/(s-1)$, and $n \ge 2$.

\noindent \textbf{Construction of Rao-Hamming Arrays}:  Construct an $s^{n} \times n$ array with the rows consisting of all $n$-tuples based on $GF(s)$. Denote the columns of this array as $C_1, \ldots, C_n$. The columns of the orthogonal array are obtained as 
$$z_1C_1 + \cdots + z_nC_n = [ C_1 \ldots C_n]z,$$
\noindent 
using all $n$-tuples $z = (z_1, \ldots, z_n)^T$ based on $GF(s)$, with $z_i \neq 0$ for at least one $i$ and the first non-zero $z_i$ equal to the unit element 1.  There are $(s^n - 1)/(s-1)$ such $n$-tuples, resulting in the required number of columns.  Theorem~\ref{thm:raohamming} summarizes the result of this construction. 
\begin{thm}\label{thm:raohamming}
If $s$ is a prime power, then an $OA(s^n, s^k,2)$  with $k = (s^n - 1)/(s-1)$ exists whenever $n \geq 2$. 
\end{thm}

The Rao-Hamming arrays achieve equality in the Rao's bound and can also be presented as linear orthogonal arrays. The number $(s^n-1)/(s-1)$ corresponds to the number of points in the finite projective geometry $PG(n-1,s)$, and the $n$-tuples used in the proof can be thought of as the points in this geometry.

\subsection{Recursive Constructions}

Recursive methods are essential tools for constructing orthogonal arrays. In Subsection~\ref{subs:difference}, we presented a recursive approach based on difference schemes. This subsection examines a recent recursive technique proposed by \cite{he2022new} for constructing larger orthogonal arrays from smaller ones with a large number of factors. 

We first define an operator based on the Kronecker product. For an $n_1 \times m_1$ matrix $A$ and $n_2 \times m_2$ matrix $B$, $n_1 \leq n_2$, and the rows of $B$ partitioned into submatrices $B_1, ..., B_{n_1}$, we define the {\em generalized Kronecker product} as 
\begin{equation}\label{eq:gks}
A \oast B = A \oast \left( \begin{array}{c}
     B_1  \\
     \vdots \\
     B_{n_1}
\end{array}\right) = [a_{i} \otimes B_{i}]_{1 \leq i \leq n_1}=
  \left(
  \begin{array}{c}
    a_{1} \otimes B_{1} \\
    \vdots \\
    a_{n_1} \otimes B_{n_1}
  \end{array}
\right),
\end{equation}
where $a_i$ denotes the $i$th row of $A$ and the operator $\otimes$ represents the usual Kronecker product defined in Subsection~\ref{subs:difference}. 
%given in (\ref{eq:ks}).  
Note that $A \oast B$ depends on the partition of the rows of $B$, that $A \oast B$ has the same number of runs as $B$, and that the number of factors in $A \oast B$ is $m_1m_2$. For example, with addition modulo 3, let $A = (0,1,2)^T$ and 
\begin{center}
$B = \left( 
\begin{array}{r}
B_1\\
B_2\\
B_3
\end{array}
\right)  =\left( \begin{array}{rrrr}
0 & 0 & 0 & 0 \\
0 & 1 & 1 & 2\\
0 & 2 & 2 & 1\\
\hdashline 
1 & 0 & 1 & 1\\
1 & 1 & 2 & 0 \\
1 & 2 & 0 & 2 \\
\hdashline 
2 & 0 & 2 & 2 \\
2 & 1 & 0 & 1\\
2 & 2 & 1 & 0
\end{array} 
\right)$, then we have 
$A \oast B =  \left( \begin{array}{rrrr}
0 & 0 & 0 & 0 \\
0 & 1 & 1 & 2\\
0 & 2 & 2 & 1\\
2 & 1 & 2 & 2\\
2 & 2 & 0 & 1 \\
2 & 0 & 1 & 0 \\
1 & 2 & 1 & 1 \\
1 & 0 & 2 & 0\\
1 & 1 & 0 & 2
\end{array} 
\right)$.
\end{center}

The construction in \cite{he2022new} uses an $n_1 \times m_1$ matrix $A$ with rows $a_1,\ldots,a_{n_1}$ and $n_1$ matrices $B_1, \ldots, B_{n_1}$ each of size $(n_2/n_1) \times m_2$ that partition the rows of $B$, with entries for all matrices from the Galois field $GF(s) =\{ \alpha_0=0, \alpha_1, \ldots, \alpha_{s-1}\}$ for a prime power $s$. They define $s+1$ arrays $D_1, D_2, \ldots, D_{s+1}$ as follows, where the Kronecker products are based on addition in $GF(s)$:
\begin{itemize} 
\item[(i)] For $\alpha_g \in GF(s)$  and $g=1,\ldots,s-1$, define 
%\begin{equation}\label{eq:construction2}
\begin{equation*}
D_{g}  =  A \oast ( \alpha_g * B) = \left( \begin{array}{c}
                                                a_1 \otimes  ( \alpha_g * B_1)\\
                                                \vdots \\
                                                a_{n_1} \otimes  ( \alpha_g * B_{n_1})
                                              \end{array}\right),
\end{equation*}
where $*$ represents multiplication in $GF(s)$.
\item[(ii)] For $g=s$, 
\begin{equation*}\label{eq:construction2s}
D_{s}  =  \bzero_{n_1} \oast   B = \left( \begin{array}{c}
                                                0 \otimes     B_1\\
                                                \vdots \\
                                                0 \otimes  B_{n_1}
\end{array}\right),
\end{equation*}
where 0 denotes the zero element in $GF(s)$ and $\bzero_{n_1}$ is the $n_1 \times 1$ vector of 0's.
\item[(iii)] For $g=s+1$, 
\begin{equation*}\label{eq:construction2splus1}
D_{s+1} = A \otimes \bzero_{(n_2/n_1)},
\end{equation*}
where $\bzero_{(n_2/n_1)}$ is the $(n_2/n_1) \times 1$ vector of 0's.
\end{itemize}
\noindent 
Now define the array
\begin{equation}\label{eq:E}
E=
    \left [D_{1},D_2, \ldots,  D_{s+1} \right ].
\end{equation}
For $1 \le g \le s-1$, $D_g$ is an $n_2 \times (m_1m_2)$, while $D_{s}$ and $D_{s+1}$ are $n_2 \times m_2$ and $n_2 \times m_1$ arrays, respectively. Consequently, $E$ is an $n_2 \times [(s-1)m_1m_2+m_1+m_2]$ array.  

The key result in  \cite{he2022new} is summarized in the following theorem. 

\begin{thm}\label{thm:e}
For a prime power $s$, if $A$ is an $OA(n_1,s^1,1)$ or an $OA(n_1,s^{m_1},2)$ for $m_1>1$, and each $B_{i}$, $i=1,\ldots,n_1$, is an $OA((n_2/n_1),s^{m_2},2)$, then the array $E$ in (\ref{eq:E}) is an $OA(n_2,s^k,2)$, where $k = (s-1)m_1m_2+m_1+m_2$.
\end{thm}

Thus, for a prime power $s$, if $A$ and the $B_{i}$'s in this construction are $s$-level orthogonal arrays of strength two, then so is the resulting array $E$. This construction uncovers the hidden structure of many existing fixed-level orthogonal arrays, often producing arrays with more factors then previously known for a given run size. 
%It also facilitates the creation of fixed-level orthogonal arrays of nearly strength three because construction methods for such arrays are limited.  
In addition, \cite{he2022new} explored how the construction in (\ref{eq:E}) can be leveraged in the construction of orthogonal arrays of strength three, as well as other types of orthogonal arrays, such as resolvable orthogonal arrays, balanced sliced orthogonal arrays, and nested orthogonal arrays. These variants of orthogonal arrays, which will be further discussed in Section~\ref{sec:recent}, offer diverse applications in experimental design and enhance flexibility for specific design goals.

\section{Distinguishing Between Orthogonal Arrays With the Same Parameters
}

Two orthogonal arrays $OA(N,s^k,t)$ with the same values for the parameters $N$, $k$, $s$ and $t$ may have different properties with respect to statistically meaningful criteria when using the orthogonal arrays as fractional factorials. Often, if there is an orthogonal array for a set of parameters, there are many such arrays, leading to the question of which one to use in an experiment. Some are isomorphic and others are non-isomorphic. Two orthogonal arrays are (algebraically) isomorphic if one can be obtained from the other by row permutations, column permutations, and/or relabeling the levels of one or more factors. Two isomorphic orthogonal arrays have similar properties with respect to some statistical criteria, but not with respect to others. Nevertheless, studying isomorphism classes is valuable for the enumeration of all possible orthogonal arrays for a given set of parameters. This is a challenging combinatorial problem due to the huge number of possible arrays. For any given array, there are after all $N! \times k! \times (s!)^k$ permutations to consider, although this number can be reduced since many of these permutations will result in the same array. With increasing values for $N$ and $k$, determining the isomorphism classes is however an NP-hard problem.

The study of isomorphism of orthogonal arrays has resulted in an extensive literature. \cite{draper1967construction} compared word-length patterns of designs to determine their isomorphism. However, two designs with the same word-length pattern could be non-isomorphic. \cite{draper1970construction} proposed a more sensitive test for isomorphism using a ``letter pattern comparison.'' However, \cite{chen1991identity} gave two non-isomorphic  $2^{31-15}$ designs with identical letter pattern matrices and thus demonstrated that 
the letter pattern also does not uniquely determine a fractional factorial.  \cite{lin2008isomorphism} subsequently showed that there are many such cases. \cite{chen1992some} discussed the isomorphism of $2^{m-p}$ fractional factorials in terms of the existence of a relabeling map between two frequency vectors together with an appropriately defined matrix. Using this frequency representation, \cite{chen1992some} proved that the word-length pattern uniquely determines any  $2^{m-p}$ fractional factorial with $p=1$ or two and any minimum aberration   $2^{m-p}$ fractional factorial when $p=3$ or 4. 
%The unpublished work \cite{chen1993catalogue} proposed a sequential algorithm for constructing complete sets of fractional factorial designs by exploring the algebraic structure of these designs. 
%A catalog of fractional factorial designs with 16, 32 and 64 runs was given. 
\cite{clark2001equivalence} introduced a method of determining isomorphism of any two (regular and non-regular) fractional factorials  by examining the Hamming distance matrices of their projection designs. An algorithm was provided for checking the isomorphism of fractional factorials when all the factors have two levels which saves considerable time for detecting non-isomorphic arrays. Other work on determining isomorphism of orthogonal arrays focuses on using different properties of the arrays such as  the centered $L_2$ discrepancy 
\citep{ma2001isomorphism}, minimal column base \citep{sun2002algorithm}, singular value decomposition \citep{katsaounis2013equivalence}, and degree of isomorphism \citep{weng2023degree}. 
\cite{katsaounis2008survey} provided a survey and evaluation of methods for determination of isomorphism of fractional factorials. 

While the criteria mentioned in the previous paragraph can be helpful in deciding whether two arrays are isomorphic, they don't solve the enumeration problem of non-isomorphic orthogonal arrays for given parameters. Nonetheless, significant progress has been made on this enumeration problem. The foundational work of \cite{chen1993catalogue} provided a comprehensive catalog of all possible  $2^{m-p}$ fractional factorials of size 16 and 32 and all resolution four (or higher) fractions of size 64.   \cite{block2005resolution} extended the enumeration to two-level 128-run resolution IV designs, offering an expanded set of designs for experimental applications requiring a larger number of runs. \cite{xu2005catalogue} proposed methods based on coding theory to efficiently classify and rank fractional factorials. This approach facilitated the enumeration of fractional factorials with 27, 81, and 243 runs at resolution IV or higher, as well as 729-run arrays at resolution V or higher. \cite{stufken2007complete} 
provided a complete solution to enumerating non-isomorphic two-level orthogonal arrays of strength $d$ with $d+2$ constraints for any $d$ and any run size $N=\lambda 2^d$. Their work represents a significant milestone, as it systematically classified all structurally unique designs within this class. \cite{schoen2010complete} introduced a minimum complete set algorithm for generating catalogs of non-isomorphic symmetric and mixed-level orthogonal arrays with specified strength, run sizes, number of factors, and number of factor levels. The cases include all mixed-level strength two orthogonal arrays with the run size $N \leq 28$, all symmetric strength two orthogonal arrays with $N 
\leq 27$, all $OA(28,2^a,2)$ with $a 
\leq 7$, all strength three orthogonal arrays with $N 
\leq 64$ except $OA(56, 2^a,3)$, $OA(64,2^a,3)$, and $OA(64,4^12^a,3)$, and all strength four orthogonal arrays with $N 
\leq 168$ except $OA(160,2^a,4)$. For $OA(56, 2^a,3)$, $OA(64,2^a,3)$, $OA(64,4^12^a,3)$ and   $OA(160,2^a,4)$, the catalog of non-isomorphic designs are obtained for $a \leq 8,7,6,8$, respectively. The accompanying Python-based software provides an extensive collection of catalogs containing non-isomorphic orthogonal arrays covering more cases than the results provided by \cite{schoen2010complete}. More recent studies, such as \cite{vazquez2019construction, bohyn2023enumeration, eendebak2023systematic}, have focused on constructing non-isomorphic orthogonal arrays for specific levels and larger run sizes. 

Another critical problem in the study of orthogonal arrays is ranking arrays with specific parameters to identify the optimal ones according to a meaningful criterion. Early foundational works include \cite{box19612} and \cite{fries1980minimum}, which introduced the optimality criteria of {\em resolution} and {\em minimum aberration}, respectively, for regular fractional factorial designs. This line of research gained momentum in the late 20th century, and numerous studies have been published since that time. Proposed optimality criteria include minimum $G$-aberration \citep{deng1999generalized}, minimum $G_2$-aberration \citep{deng1999minimum}, minimum moment aberration \citep{xu2003minimum}, various uniformity measures \citep{fang2000miscellanea}, and estimation capacity \citep{cheng1999minimum}. For a comprehensive review, see \cite{xu2009recent} and \cite{cheng2016theory}.

\section{Selected Recent Developments} \label{sec:recent}

In recent years, practical applications in science, engineering, and technology have led to multiple innovations in research related to orthogonal arrays. This section highlights some of the most notable developments that have drawn particular interest within the experimental design community.
These include sliced orthogonal arrays, nested orthogonal arrays, strong orthogonal arrays, and grouped orthogonal arrays. The following subsections provide an overview of these concepts and key advancements in their development. 

\subsection{Sliced Orthogonal Arrays}

Sliced orthogonal arrays were developed to address the needs of computer experiments involving both quantitative and qualitative input variables, a scenario that frequently occurred in real-world applications. For example, a data center experiment might include qualitative factors such as diffuser height and the location of hot-air return vents \citep{schmidt2005challenges}. Similarly, computer experiments in marketing and social sciences often involve qualitative factors such as education level, race, and social background. \cite{qian2009sliced} was the first to introduce the concept of sliced orthogonal arrays to tackle experimental design challenges in such mixed-input computer experiments.

To define sliced orthogonal arrays, we first define the concept of a {\em level-collapsing projection}.  A mapping $\delta(\cdot)$ is called a level-collapsing projection if the mapping is from a set $S$ with $s$ elements to a set of $s_0$ elements and satisfies (a) $S$ can be partitioned into $s_0$ subsets $S_1, \ldots, S_{s_0}$, with each having $s/s_0$ elements, and (b) for any two elements $x\in S_i, y\in S_j$, $\delta(x)=\delta(y)$ for $i=j$, and $\delta(x)\neq\delta(y)$ otherwise.  Table~\ref{tab:OAs} provides an example of the level-collapsing projection $\delta(0)=\delta(1) = 0$ and $\delta(2) = \delta(3) = 1$. 

\begin{defn}
An $OA(N,s^k,2)$ $D$ is called a sliced orthogonal array if the $N$ rows of $D$ can be partitioned into $v$ subarrays $D_1, D_2, \ldots, D_v$  such that   each $D_i$ becomes an $OA(N_0,s_0^k,2)$ with $N_0 = N/v$ after the $s$ levels in each column of $D$ are collapsed to $s_0$ levels according to some level-collapsing projection.
\end{defn}

We denote such an array by $SOA(N,s^k,2;v,s_0)$. Furthermore, if each column in each slice $D_i$ of a sliced orthogonal array is balanced, that is, has an equal occurrence of each of the $s$ levels, it is called a {\em balanced sliced orthogonal array} \citep{ai2014construction}. We use $BSOA(N,s^k,2;v,s_0)$ to denote such an array. The sliced orthogonal array $D$ in Table~\ref{tab:OAs} is a $BSOA(16,4^3,2;4,2)$, where rows 1--4, 5--8, 9--12, 13--16 correspond to $D_1$, $D_2$, $D_3$ and $D_4$, respectively.  

\begin{table}[!ht] \begin{center}
\caption{An orthogonal array $D=OA(16,4^3,2)$ and the level-collapsing projection $\delta$: $D$   is also  a $BSOA(16,4^3,2;4,2)$ and an $NOA(16,4^3,2;4,2)$} 
\scalebox{0.9}{
\begin{tabular}{cccccccc}
\multicolumn{8}{c}{}\\
                                 \multicolumn{3}{c}{$D$}  &  &  &   \multicolumn{3}{c}{$\delta(D)$}  \\ \hline                          
          $0$     &   $0$      &     $0$   &  &  &  $0$     &   $0$      &     $0$    \\
                                    $2$     &   $1$      &    $3$  &  &  &  $1$     &   $0$      &    $1$     \\
                                    $1$     &   $3$    &     $2$   &  &  &  $0$     &   $1$      &     $1$    \\
                                  $3$   &   $2$      &    $1$    &  &  &  $1$     &   $1$      &    $0$     \\ 
                                  %\hdashline 
         $2$     &   $2$      &     $2$   &  &  &  $1$     &   $1$      &     $1$    \\
                                    $0$     &   $3$    &    $1$    &  &  &  $0$     &   $1$      &    $0$     \\
                                    $3$   &   $1$      &     $0$   &  &  &  $1$     &   $0$      &     $0$    \\
                                    $1$     &   $0$      &    $3$  &  &  &  $0$     &   $0$      &    $1$    \\ %\hdashline 
  $1$     &   $1$      &     $1$   &  &  &  $0$     &   $0$      &     $0$    \\
                                   $3$   &   $0$      &    $2$    &  &  &  $1$     &   $0$      &    $1$     \\
                                    $0$     &   $2$      &    $3$  &  &  &  $0$     &   $1$      &    $1$     \\
                                    $2$     &   $3$    &    $0$    &  &  &  $1$     &   $1$      &    $0$     \\ % \hdashline
    $3$   &    $3$   &   $3$   &  &  &  $1$     &   $1$      &   $1$      \\
                                    $1$     &   $2$      &    $0$    &  &  &  $0$     &   $1$      &    $0$     \\
                                    $2$     &   $0$      &     $1$   &  &  &  $1$     &   $0$      &     $0$    \\
                                    $0$     &   $1$      &    $2$    &  &  &  $0$     &   $0$      &    $1$     \\
\end{tabular}}\label{tab:OAs}
\end{center}
\end{table}

A sliced orthogonal array can be used to construct sliced space-filling designs, which are used to choose inputs for computer experiments with both qualitative and quantitative inputs. Suppose we have $v$ level combinations of the qualitative factors, and $k$ quantitative input variables in the experiment. We will use a sliced orthogonal array with $v$ slices and $k$ factors. The procedure of constructing sliced space-filling designs consists of  two main steps. First, for the quantitative factors,  a Latin hypercube design is generated using a sliced orthogonal array
in the same way as done for generating an orthogonal array-based Latin hypercube (see Subsection~\ref{subs:3.1}), with the rows of the resulting Latin hypercube designs being partitioned into $v$ slices corresponding to the partition in the sliced orthogonal array.  Second, 
each of these $v$ slices is then associated with a level combination of the qualitative factors.   
This approach ensures an efficient and balanced design for experiments with mixed inputs. For example. the sliced orthogonal array in Table~\ref{tab:OAs} can be used to generate sliced space-filling designs for a computer experiment with 
three quantitative input variables and four level combinations for qualitative factors, where the four level combinations could form a full factorial for two qualitative factors or a fractional factorial for three qualitative factors. 

 \cite{qian2009sliced} introduced several methods for constructing sliced orthogonal arrays, including based on the Rao-Hamming method. Building on this pioneering work, additional construction methods have been developed, as detailed in, among others, \cite{xu2011sudoku, ai2014construction, li2015sliced, hwang2016sliced, zhang2018some, he2019sliced, tsai2022some, he2022new,pang2024construction}, to construct sliced orthogonal arrays with more flexible parameters.

\subsection{Nested Orthogonal Arrays}

\cite{qian2009nested} introduced the concept of {\em nested orthogonal arrays} to construct nested space-filling designs, which are used to select inputs for computer experiments at two levels of accuracy or fidelity. Such experiments are commonly seen in practice when computer models can be run with varying degrees of sophistication, resulting in different computational times.
For instance, \cite{qian2008bayesian} studied computer simulations for a heat exchanger in an electronic cooling application. In this case, two computer codes were employed to simulate linear cellular alloys for electronic cooling systems. One code used finite-element analysis, providing high accuracy but requiring more computational time, while the other relied on the finite difference method, yielding a faster but less precise approximation. The differences in numerical methods and grid resolution created a trade-off between accuracy and computational efficiency, highlighting the need for nested designs.

%A nested orthogonal array is defined as follows. 

\begin{defn}
An $OA(N,s^k,2)$ $D$  is called a nested orthogonal array if 
$D$ contains a subarray  $D_0$ that, after applying a specific level-collapsing projection to each column of $D$, becomes an $OA(N_0,s_0^k,2)$. 
\end{defn}

We denote such an array by $NOA(N,s^k,2;N_0,s_0)$.
For illustration, consider the $BSOA\\(16,3,4,2;4,2)$ $D$ in Table~\ref{tab:OAs}. This array is also an $NOA(16,4^3,2;4,2)$  where any 
 $D_i$ for $i=1, 2, 3, 4$, can serve as the subarray  $D_0$ (rows 1--4, 5--8, 9--12, 13--16 correspond to $D_1$, $D_2$, $D_3$ and $D_4$, respectively). In fact, a sliced orthogonal array is a nested orthogonal array, however the reverse does not hold. 

\cite{mukerjee2008existence} and \cite{wang2013further} explored the existence of nested orthogonal arrays. Building on the work of \cite{qian2009nested}, many studies have investigated methods for constructing nested orthogonal arrays, including  \cite{qian2009construction, dey2010construction, he2011nested, dey2012construction, wang2013constructions, sun2014construction, zhang2018some, zhang2019construction, he2022new, pang2024construction}.
In general, nested orthogonal arrays can be constructed through various direct methods, such as the Rao-Hamming method, Bush's method, or by leveraging structures like nested difference matrices, Hadamard matrices, resolvable orthogonal arrays, zero-sum arrays, and operators like the Kronecker product and subgroup projection. Recursive methods are also used, offering an iterative approach to generate larger nested orthogonal arrays from smaller ones.
These construction methods aim to provide greater flexibility in terms of run sizes and the number of levels for each factor, enabling the development of nested designs tailored to a wide range of experimental requirements.

\subsection{Strong Orthogonal Arrays}

\cite{he2013strong} introduced the concept of {\em strong orthogonal arrays}, which are used to generate space-filling designs with enhanced projection and space-filling properties. These arrays have broad applications in optimization, prediction, and sensitivity analysis of complex systems, where effective exploration of the design space is critical. The introduction of strong orthogonal arrays by \cite{he2013strong} has received considerable interest from both researchers and practitioners, leading to advancements in both design theory and practical applications.

\begin{defn} \label{def:strong}
An $N \times k$ array with entries from \{$0,1,\ldots, s^t-1$\} is called a strong orthogonal array of size $N$, $k$ factors, $s^t$ levels, and strength $t$ if any subarray of $g$ columns for any $g$ with $1\leq g \leq t$ can be collapsed into an $OA(N,s^{u_1} s^{u_2} \cdots s^{u_g},g)$ for any positive integers $u_1, \ldots, u_g$ with $u_1 + u_2  + \cdots + u_g = t$ where collapsing $s^t$ levels into $s^{u_j}$ levels is done through the map $a \rightarrow [a/s^{t-u_j}]$, with $[x]$ denoting the largest integer not exceeding $x$. 
\end{defn}

Note that in Definition~\ref{def:strong}, the notation $s^{u_1}$ stands for a single factor at $s^{u_1}$ levels rather than for $u_1$ factors at $s$ levels. Similarly for the other $g-1$ factors.
We use $SOA(N,(s^t)^k, t)$ to denote such a strong orthogonal array. Note that, despite sharing the same acronym as sliced orthogonal arrays, the notation differs in both the type and number of parameters. Table~\ref{table:soa} presents an $SOA(8, 8^3, 3)$. This array has the following properties:\\[-0.3in]
\begin{itemize}
\item[(i)] The array becomes an $OA(8,2^3,3)$ after the eight levels are collapsed into two levels according to $[a/4] = 0$ for $a=0,1,2,3$ and $[a/4] = 1$ for $a = 4,5,6,7$, where $[x]$ denotes the largest integer not exceeding $x$;
\item[(ii)] Any two columns of the array become an $OA(8, 2^14^1,2)$ or $OA(8,4^12^1,2)$ after the eight levels of one column are collapsed into two levels by $[a/4]$ and the eight levels of the other column are collapsed into four levels by $[a/2]$;
\item[(iii)] Any column of the array is an $OA(8,8^1,1)$.     
\end{itemize}

\begin{table}[!h]
\begin{center}
\caption{An $SOA(8,8^3,3)$}
\begin{tabular}{ccc}
\multicolumn{3}{c}{}\\
 0 & 0 & 0 \\
 2 & 3 & 6\\
 3 & 6 & 2 \\
 1 & 5 & 4\\
 6 & 2 & 3 \\
 4 & 1 & 5\\
 5 & 4 & 1\\
 7 & 7 & 7 
\end{tabular}\label{table:soa} 
\end{center}
\end{table}

Clearly, $N = \lambda s^t$ must hold for some integer $\lambda$. As was the case for an orthogonal array, $\lambda$ is called the index of the strong orthogonal array. \cite{he2013strong} noted that the definition of a strong orthogonal array is motivated by nets for quasi-Monte Carlo point sets, and nets are a special case of strong orthogonal arrays. Their connection is documented in \cite{he2013strong}. If $\lambda = s^w$ for some integer $w$, then the existence of an $SOA(\lambda s^t, (s^t)^k,t)$ is equivalent to that of a $(w,m,k)$-net with base $s$ where $m=w+t$.  Nets are defined with the restriction that the index is a power of $s$ while strong orthogonal arrays do not have such a restriction. 

Looking into the definition of strong orthogonal arrays, if we focus on $t=2$, we can see that an $SOA(N,(s^2)^k,2)$ of strength two is an orthogonal array of strength one itself and becomes an orthogonal array of strength two if its $s^2$ levels are collapsed into $s$ levels according to $[a/s]$. This implies that an $SOA(N,(s^2)^k,2)$ has the same projection property as an orthogonal array of strength two. An $SOA(N,(s^3)^k,3)$ however, offers better stratification and projection property than an $s$-level orthogonal array of strength three. This is because an $SOA(N,(s^3)^k,3)$ achieves stratification on $s^2 \times s$ and $s \times s^2$ grids in two-dimensions and $s \times s \times s $ grids in three-dimensions while an $s$-level orthogonal array of strength three can only promise stratification on $s \times s $ grids in two-dimensions and $s \times s \times s $ grids in three-dimensions. Similar examinations reveal that to enjoy the benefits of better space-filling properties, when compared to ordinary orthogonal arrays, strong orthogonal arrays need to have strength three or higher which may require run sizes that are too large for experimenters to afford in practice. 
\cite{he2018strong} introduced
a new class of arrays, called strong orthogonal arrays of strength two plus.

\begin{defn}
An $N \times k$ array with entries from \{$0,1,\ldots, s^2-1$\} is called a strong orthogonal array of size $N$, $k$ factors, $s^2$ levels, and strength $2+$ if any subarray of two columns can be collapsed into an $OA(N,(s^2)^1s^1,2)$ and an $OA(N,s^1(s^2)^1,2)$.
\end{defn}

Since their introduction, several research topics on strong orthogonal arrays   have been explored, leading to significant advancements in their theory and applications. 
First, a number of studies have focused on developing construction methods for strong orthogonal arrays, particularly those of strength three and strength two plus. Notable contributions in this area include \cite{he2018strong} and \cite{shi2020construction}. Second, selecting an optimal strong orthogonal array from the broader class based on specific design criteria remains a fundamental problem. Representative works addressing this challenge include \cite{shi2019design} and \cite{chen2024selecting}. Understanding and characterizing strong orthogonal arrays is another critical research direction. \cite{he2014characterization} made significant contributions in this area, providing deeper insights into the properties of strong orthogonal arrays. In addition, substantial advancements have been achieved by imposing additional structures on strong orthogonal arrays. For example, \cite{li2021column} and \cite{zhou2019column} considered column-orthogonal strong orthogonal arrays; \cite{liu2015column} studies sliced strong orthogonal arrays;  \cite{zheng2024nested} introduced nested strong orthogonal arrays; \cite{wang2022construction} investigated strong group-orthogonal arrays. \cite{shi2023evaluating} empirically showed the advantage of strong orthogonal arrays in  hyperparameter tuning in deep neural networks, providing an example of the use of strong orthogonal arrays in machine learning. 
 
\subsection{Grouped Orthogonal Arrays}

\cite{GOA2024} introduced the concept of grouped orthogonal arrays, motivated by applications in computer experiments where interactions between factors only arise from disjoint groups of variables. In such experiments, the true response surface or the preferred surrogate model is additive, with each component being a function of one specific group of variables. A practical example is provided in the engine block and head joint sealing experiment discussed by \cite{joseph2008blind}, where eight factors were involved. Their analysis identified significant linear, quadratic, and interaction effects among the first, second, and sixth factors. The eight factors can be divided into two groups: one comprising the first, second, and sixth factors, and the other containing the remaining factors. In such cases, a design with superior projection and space-filling properties for the first group is more desirable than traditional space-filling designs that lack this feature because it enables more accurate estimation of the main and interaction effects within the first group. 

\begin{defn}
An orthogonal array is called an $s$-level grouped orthogonal array with $N$ runs, $g$ groups  and strength $t$ if its factors can be partitioned into $g$ groups where the $i$th group has $k_i$ factors and is of strength $t_i$, where $t_i \geq t$, for $i =1, \ldots, g$.   
\end{defn} 
We use $GOA(N, (k_1,k_2,\ldots, k_g),  (t_1,t_2, \ldots, t_g), s, t)$ to denote such an array.  An an example, Table~\ref{tab:d27} displays a $GOA(27; (4; 3; 3); (3,3,3); 3; 2)$, for which the whole array is a three-level orthogonal array of strength two, but where the columns are partitioned into three groups, denoted by $D_1$, $D_2$ and $D_3$ in Table~\ref{tab:d27}, that each form an orthogonal array of strength three. 
\begin{table}[!ht] \begin{center}
\caption{The design matrix $D=(D_1,D_2,D_3)$ for GOA$(27, (4,3,3), 3 \times 3, 3, 2)$.}
\centering
\begin{tabular}{ccc}
  \hline
$D_1$ & $D_2$ & $D_3$ \\
  \hline
{0000}  & {000}  & {000}  \\
{1110}  & {111}  & {111}  \\
{2220}  & {222}  & {222}  \\
{0120}  & {012}  & {012}  \\
{1200}  & {120}  & {120}  \\
{2010}  & {201}  & {201}  \\
{0210}  & {021}  & {021}  \\
{1020}  & {102}  & {102}  \\
{2100}  & {210}  & {210}  \\
{0111}  & {122}  & {200}  \\
{1221}  & {200}  & {011}  \\
{2001}  & {011}  & {122}  \\
{0201}  & {101}  & {212}  \\
{1011}  & {212}  & {020}  \\
{2121}  & {020}  & {101}  \\
{0021}  & {110}  & {221}  \\
{1101}  & {221}  & {002}  \\
{2211}  & {002}  & {110}  \\
{0222}  & {211}  & {100}  \\
{1002}  & {022}  & {211}  \\
{2112}  & {100}  & {022}  \\
{0012}  & {220}  & {112}  \\
{1122}  & {001}  & {220}  \\
{2202}  & {112}  & {001}  \\
{0102}  & {202}  & {121}  \\
{1212}  & {010}  & {202}  \\
{2022}  & {121}  & {010}  \\
  \hline
\end{tabular}\label{tab:d27}
\end{center}
\end{table}

The concept of grouped orthogonal arrays is not new. In addressing the experimental design issue in physical experiments and applications in combinatorics, \cite{lin2012designs} introduced 
designs of variable resolution and \cite{Raaphorst-Moura-Stevens-2014} coins variable strength orthogonal arrays. \cite{lin2012designs} and \cite{lekivetz2016designs} provided several constructions for designs of variable resolution but the focus was on two-level designs.  
The variable strength orthogonal arrays obtained by \cite{Raaphorst-Moura-Stevens-2014} have groups of three factors and their run sizes are limited to $s^3$ for a prime power $s$. \cite{Zhang-Pang-Li-2023} constructed variable strength orthogonal arrays with strength $l$ containing a subarray with strength greater than $l$, where $l \geq 2$. In addition to the drawback that the designs constructed have only one group with larger strength, the resulting designs have very restrictive run sizes $s^t$ for a prime power $s$ and an integer $t \geq 4$. \cite{GOA2024} proposed several construction methods to generate many more designs with 
flexible run sizes and better within-group projection properties for any prime power number of levels.

The research on grouped orthogonal arrays is quite new. Several important directions are called for. First, current constructions mostly produce regular designs and thus construction methods for non-regular grouped orthogonal arrays are needed. Second, constructions on grouped orthogonal arrays with differing group sizes and mixed-level grouped orthogonal arrays are worth exploring.   Another important topic is the use of group orthogonal arrays in analysis of computer experiments and beyond such as models with blocked additive kernels, as done in \cite{lin2014design}, which showed the advantages of designs of variable resolution in model selection of linear models.

%\subsection{Mappable nearly orthogonal arrays}

\section{Summary}

% {\red To be completed later}

This review is based on the experiences and interests of the authors, and therefore incomplete and selective. Yet, even this selective review demonstrates the enormous impact that the introduction of orthogonal arrays by \cite{rao1946hypercubes, rao1947factorial, rao1949class} has had on applications and research. Orthogonal arrays are a simple yet elegant mathematical and statistical tool with a rich theoretical foundation and diverse applications across many fields. This review highlights fundamental results and recent advancements, hoping to interest more readers in research on orthogonal arrays and related structures or to use orthogonal arrays in novel applications. In an era where, in some fields, large datasets are regularly collected, there continue to be emerging roles for orthogonal arrays, as we have pointed out in this review. This includes applications in data subsampling and machine learning. There is no doubt that orthogonal arrays will remain a prominent and versatile tool that will continue to drive innovation, including in the rapidly evolving realm of artificial intelligence. 

\section*{Conflicts of Interest Statement}
The authors have no conflicts of interest to disclose.
%\begin{figure}[bt]
%\centering
%\includegraphics[width=6cm]{example-image-rectangle}
%\caption{Although we encourage authors to send us the highest-quality figures possible, for peer-review purposes we can accept a wide variety of formats, sizes, and resolutions. Legends should be concise but comprehensive -- the figure and its legend must be understandable without reference to the text. Include definitions of any symbols used and define/explain all abbreviations and units of measurement.}
%\end{figure}

%

%\section*{Supporting Information}
%{\red To be completed later}

%Supporting information is information that is not essential to the article, but provides greater depth and background. It is hosted online and appears without editing or typesetting. It may include tables, figures, videos, datasets, etc. More information can be found in the journal's author guidelines or at \url{http://www.wileyauthors.com/suppinfoFAQs}. Note: if data, scripts, or other artefacts used to generate the analyses presented in the paper are available via a publicly available data repository, authors should include a reference to the location of the material within their paper.

%\printendnotes

\bibliographystyle{chicago}   
\bibliography{OAreview} % Add your references here

\end{document}